\documentclass[useAMS,amssymb,usenatbib]{mn2e}

\usepackage{epsfig,natbib,deluxetable}

\title[]{Globular Cluster Systems in Nearby Dwarf Galaxies: \\
I. HST/ACS Observations and Dynamical Properties of Globular 
Clusters at Low Environmental Density}

\author[I.\,Y.\,Georgiev et al.]{Iskren Y. Georgiev$^{1,2}$\thanks{E-mail: iskren@astro.uni-bonn.de}, Thomas H. Puzia$^{3}$, Michael Hilker$^{4}$, and Paul Goudfrooij$^{2}$\\
$^{1}$Argelander Institut f\"{u}r Astronomie der Universit\"{a}t Bonn, Auf dem H\"{u}gel 71, D-53121 Bonn, Germany\\
$^{2}$Space Telescope Science Institute, 3700 San Martin Drive, Baltimore, MD 21218, USA\\
$^{3}$Herzberg Institute for Astrophysics, 5071 Saanich Road, Victoria, BC V9E 2E7, Canada\\
$^{4}$European Southern Observatory, D-85748 Garching bei M\"unchen, Germany
}
\begin{document}

\date{October 2008}

\pagerange{\pageref{firstpage}--\pageref{lastpage}} \pubyear{2008}

\maketitle

\label{firstpage}

\begin{abstract}
We investigate the old globular cluster (GC) population of 68 faint
($M_V>-16$ mag) dwarf galaxies located in the halo regions of nearby 
($\la12$ Mpc) loose galaxy groups and in the field environment based
on archival HST/ACS images in F606W and F814W filters. The combined color
distribution of 175 GC candidates peaks at $(V-I)=0.96\pm0.07$ mag and the
GC luminosity function turnover for the entire sample is found at
$M_{V,{\rm TO}}\!=\!-7.6\pm0.11$ mag, similar to the old metal-poor LMC GC
population. Our data reveal a tentative trend of $M_{V,{\rm TO}}$ becoming
fainter from late-type to early-type galaxies. The luminosity and color
distributions of GCs in dIrrs shows a lack of faint blue GCs. Our analysis
reveals that this might reflect a relatively younger GC system than 
typically found in luminous early-type galaxies. If verified by 
spectroscopy this would suggest a later formation epoch of the first 
metal-poor star clusters in dwarf galaxies. We find several bright 
(massive) GCs which reside in the nuclear regions of their host 
galaxies. These nuclear clusters have similar luminosities and 
structural parameters as the peculiar Galactic clusters suspected of 
being the remnant nuclei of accreted dwarf galaxies, such as M54 and
$\omega$Cen. Except for these nuclear clusters, the distribution of GCs in
dIrrs in the half-light radius vs. cluster mass plane is very similar to
that of Galactic young halo clusters, which suggests comparable formation
and dynamical evolution histories. A comparison with theoretical models of
cluster disruption indicates that GCs in low-mass galaxies evolve
dynamically as self-gravitating systems in a benign tidal environment.
\end{abstract}


\section{Introduction}\label{intro}
Globular clusters (GCs) are observed in vast numbers in massive early-type
galaxies \cite[e.g.][]{Harris06a} and the integrated-light properties of GCs were
extensively studied in the last decade \cite[e.g.][]{Kissler-Patig97, Hilker99,
  Kundu&Whitmore01, Kundu&Whitmore01b, Larsen01, Goudfrooij03,
  Puzia02, Puzia04, Harris06b, Jordan05, Jordan07, Peng06, Peng08}
with the aim to understand how such populous GC systems were assembled. These
studies revealed the discovery of {\it i}) the bimodal
metallicity/color distribution of GCs \cite[e.g.][and references
therein]{Ashman&Zepf92, Gebhardt&Kissler-Patig99,  Puzia99, Kundu&Whitmore01} and
{\it ii}) the presence of young and intermediate-age globular clusters in merging,
starburst, and quiescent galaxies \cite[e.g.][]{Whitmore&Schweizer95,
  Goudfrooij01, Goudfrooij07, Puzia02, Schweizer02}. 

From the point of view of GC system assembly, multiple scenarios for galaxy
formation have been presented to explain these findings. The first
scenario is the {\it hierarchical\/} build-up of massive galaxies from
pregalactic dwarf-sized gas fragments \cite[]{Searl&Zinn78} in which the
metal-poor GCs form {\it in situ\/} while the metal-rich GCs originate from
a second major star formation event \cite[]{Forbes97} or infalling gaseous
fragments, the so-called mini-mergers at high redshifts
\cite[]{Beasley02}. The {\it dissipative merger\/} scenario
\cite[e.g.][]{Schweizer87, Ashman&Zepf92, Bekki02} assumes that the
metal-poor GCs were formed early in ``Searle-Zinn'' fragments while
metal-rich (and younger) GCs formed during major merger events of galaxies
with comparable masses (Spiral--Spiral, Elliptical--Spiral, etc.). The
{\it dissipationless merging and accretion\/} scenario incorporates the
classical monolithic galaxy collapse in which the galaxies and their GCs
were formed after which the dwarf galaxies and their GCs were accreted by
giant galaxies \cite[]{Zinn93, Cote98, Cote02, Hilker99, Lee07}. For
detailed discussions on the GC systems formation and assembly scenarios 
and properties of GCs in various galaxy types we refer the reader to 
\cite{Kissler-Patig00, vdBergh00, Harris01, Harris03, Brodie06}.

Dwarf galaxies play an important role in assessing the likelihoods of the
formation scenarios mentioned above. 
The prediction of the hierarchical growth of
massive galaxies through merging of many dwarf-sized fragments at early
times is backed by the steepening of the faint-end slope of the galaxy
luminosity function with redshift \cite[e.g.][]{Ryan07, Khochfar07}. In
addition, the observed fraction of low-mass irregular galaxies increases
with redshift \cite[e.g.][]{Conselice08}, and Hubble Ultra Deep field
studies show that dwarf-sized irregular galaxies dominate at $z\ga6$
\cite[e.g.][]{Stiavelli04}. Hence, the oldest stellar populations in
nearby dIrr galaxies may thus represent the probable {\it surviving} early
building blocks of massive galaxies.

Evidence for accretion events of dwarf galaxies and their dissolution in
the Galactic potential is manifested by a few tidal stellar 
streams observed in the Milky Way, M31, and other nearby galaxies
\cite[e.g.][and references therein]{Ibata01, Grillmair06, Liu06,
Martinez-Delgado08}. A well-known example of a recent accretion
event is the Sagittarius dwarf galaxy, which is currently being added to
the Milky Way halo together with its globular clusters
\cite[e.g.,][]{Ibata94}. Many more such minor accretion/merger events 
of dwarf sized galaxies are expected during the hierarchical evolution 
of giant galaxies. The \cite{Pipino07}
galaxy evolution and GC formation model, in which a low number of blue
(metal-poor) GCs in massive galaxies are predicted to form in the initial
("monolithic") collapse, contrasts with observations and suggests that
subsequent accretion of such metal-poor GCs from low-mass dwarfs is
required \citep{Cote98, Cote02}. However, the observed large ratio between
metal-poor GCs and metal-poor field halo stars in massive galaxies
\cite[]{Harris01, Harris&Harris02} implies that, if such dwarfs were later
accreted, they must have had large GC specific frequencies
$S_{N}$\footnote{$S_N$ is the number of GCs per unit galaxy luminosity
\citep{Harris&vdBergh81}.} in order to keep the GC-to-field-star ratio
high. 
This is in line with the high specific frequencies observed in
low-mass early and late-type dwarfs in galaxy clusters \cite[]{Durrell96,
Miller98, Seth04, Miller&Lotz07, Georgiev06, Peng08} and in group/field
environments \cite[]{Olsen04, Sharina05, Georgiev08, Puzia&Sharina08}.

In this series of papers we will present results from an analysis of old 
GCs in 68 dwarf galaxies in nearby loose groups and in the field. The current 
paper is organized as follows. In \S2 we describe the target sample, discuss 
the completeness and contamination, and define the globular cluster candidate
selection. In \S3 the colors, luminosities, and structural parameters are 
compared with old Local Group GCs, and a discussion of the dynamical state 
of the globular clusters in our sample dwarfs. Section \S4 summarizes our results. 

\section{Data}
\subsection{Observations}\label{description}

The current study is based on HST/ACS archival data from programs
SNAP-9771 and GO-10235, conducted in HST cycles 12 and 13, respectively (PI: I.
Karachentsev), and archival HST/ACS data (GO-10210, PI: B. Tully) that were 
described in \cite{Georgiev08}. A summary of the various datasets is
provided in Table~\ref{tab:sum}.

\begin{table}
 \centering
  \caption{Summary of target sample. The table shows the number of morphological
    galaxy types observed in each program and the total sample. The bottom row
    shows the number of galaxies in which we found globular cluster candidates.} 
  \begin{tabular}{@{}lrrrr@{}}
\hline\hline
 Program ID	&  dIrr	&  	dE 	& 	dSph	& 	Sm\\
\hline\hline
 SNAP~9771	&  26 	& 	2 	&	4 		& 	2 \\
 GO-10235	&  10 	& 	1 	&	1 		& 	3 \\
 GO-10210	&  19 	& 	-- 	&	-- 		& 	-- \\
\hline
$\Sigma_{\rm all}$	&  55 	& 	3 	&	5	& 	5 \\
$\Sigma_{\rm w/GC}$&  30 	& 	2 	&	2	& 	4 \\
\hline\hline\label{tab:sum}
\end{tabular}
\end{table}

Non-dithered $2\times600$\,s F606W and $2\times450$\,s F814W exposures for each
galaxy were designed to reach the Tip of the Red Giant Branch (TRGB) at
$M_I=-4.05$ or $M_V=-2.75$ to $-1.45$\,mag in the range [Z/H]~$= -2.2$ to $-0.7$
dex \cite[][]{DaCosta&Armandroff90} and provide TRGB distances to the sample dwarf
galaxies as published in \cite{Karachentsev06, Karachentsev07}. All galaxies
reside in the isolated outskirts of nearby ($\leq6$\,Mpc) galaxy groups (Sculptor,
Maffei\,1\&\,2, IC\,342, M\,81, CVn\,I cloud) with the exception of 3 very
isolated dwarfs within 12\,Mpc. 

The program SNAP-9771 observed 34 low-mass galaxies, of which 26 are dIrrs, 2 dEs,
4 dSphs and 2 late-type spiral dwarf galaxies. The program GO-10235 contains 15
targets of which 10 are dIrrs, 1 dE, 1 dSph and 3 late-type spirals. All galaxies
in our final sample have published TRGB distance measurements
\cite[]{Karachentsev06, Karachentsev07}. A summary of their general properties is
given in Table~\ref{glist}. In column (1) the galaxy IDs are listed, and in
columns (2) and (3) their coordinates; columns (4) and (5) contain the
morphological classifications from LEDA\footnote{http://leda.univ-lyon1.fr/} and
NED\footnote{http://nedwww.ipac.caltech.edu/}, (6) and (7) list the distances and
distance moduli, in (8) is the foreground galactic extinction ($E_{B-V}$), (9) 
and (10) give the absolute magnitudes and colors (measured in this work, see 
Section~\ref{SB}), and in column (11) we provide their {\sc HI} mass (obtained 
from LEDA). 

The additional archival HST/ACS data consists of 19 Magellanic-type dIrr galaxies
residing in nearby ($2 - 8$\,Mpc) associations composed mainly of dwarfs with
similar luminosities as the target galaxies, with the only exception that the
previous sample was composed of dwarfs located in isolated associations of only
few dwarfs \cite[]{Tully06}, while the dwarfs in our new dataset are in the halo
regions of groups which contain massive dominant galaxies such as M\,83 and Cen\,A
\cite[see Table~\ref{glist}] {Karachentsev06, Karachentsev07}. The HST/ACS F606W
and F814W imaging of the previous data was also designed to reach the tip of the
red giant branch (TRGB) and measure TRGB distances \cite[]{Tully06}. 

In total, our sample consists of 55 dIrrs, 3 dEs, 5 dSphs, and 5 Sm galaxies in
the field environment. All dwarfs have apparent diameters smaller than the HST/ACS
field of view which provided us with a good sampling of their GCSs. 

\subsection{Data Reduction and Photometry}
Image processing and photometry was performed in a manner identical to that in 
our previous study described in \cite{Georgiev08}. In the following we
briefly summarize the basic steps. We retrieved archival HST/ACS images in
F606W and F814W filters which were processed with the ACS reduction
pipeline and the {\sc multidrizzle} routine \cite[]{Koekemoer02}. To
improve the object detection and photometry we modeled and subtracted
the underlying galaxy light using a circular aperture median kernel of
41\,pixel radius. This choice of filter radius at 3\,Mpc (the closest 
galaxy) corresponds to $\sim30$\,pc, which is ten times the typical GC 
$r_{\rm h}$. This is large enough that after median galaxy light subtraction, 
the structure of the GC candidates will be preserved.

Object detection ($4\sigma$ above the background) and initial aperture
photometry (in 2,3,5 and 10 pixel radius) was performed with the
IRAF\footnote{IRAF is distributed by the National Optical Astronomy
Observatories, which are operated by the Association of Universities for
Research in Astronomy, Inc., under cooperative agreement with the National
Science Foundation.} {\sc DAOPHOT/DAOFIND} and {\sc PHOT} routines.
Conversion from instrumental magnitudes to the STMAG system and aperture
corrections from 10 pixel to infinity were performed using the ACS
photometric calibration prescription by \cite{Sirianni05}.

\subsection{GC Candidate Selection}
The first step in our GC candidate\footnote{In the following we refer to globular cluster candidates
  as GCs.} selection was based on the expected colors for old
($>\!4$\,Gyr) Simple Stellar Population (SSP) models \cite[]{Anders03, BC03} and
typical colors of old Galactic GCs, namely objects in the color range
$-0.4\leq$(F606W-F814W)$\leq0.2$ (corresponding to $0.7<V-I<1.5$). Second, we have
imposed a faint magnitude cut-off at the TRGB \cite[$M_V=-2.5$,
$M_I=-4.05$;][]{DaCosta&Armandroff90,Lee93} converted to apparent magnitudes
according to the distance to each galaxy as derived by \cite{Tully06} and
\cite{Karachentsev06,Karachentsev07}. This is $\sim\!5$\,mag fainter than the
typical GC luminosity function turnover magnitude for metal-poor Galactic globular
clusters at $M_{V,\,{\rm TO}}=-7.66\pm0.1$\,mag \cite[e.g.][]{DiCriscienzo06}. 

Due to the deep ACS imaging and its high spatial resolution (1 ACS pixel being
equivalent to 1.2\,pc at a mean distance to our sample dIrrs of $5$\,Mpc), a typical
Local Group GC with half-light radius of $r_{\rm h}=3$\,pc is resolved. 
We used the {\sc imexam} task within IRAF to confine the initial GC selection to
round objects (FWHM$_{F606W}\simeq$~FWHM$_{F814W}$) with an {\sc imexam} ellipticity
$\epsilon\leq0.15$ within a fixed $r=5$\,pix aperture radius and a Moffat
index $\beta=2.5$, typical for stellar profiles. As we showed in
\cite{Georgiev08}, the upper {\sc imexam} $\epsilon$ limit allows
selection of GCs with true ellipticity up to $\epsilon<0.4$. The
final ellipticities and half-light radii ($\epsilon,\ r_{\rm h}$) were
measured with {\sc ishape} (see Sect.\,\ref{sizes}).

The last GC selection step is to measure the differences between the
aperture photometry determined using 2, 3, and 5 pixel radii. These m$_{2}-$m$_3$
vs. m$_{2}-$m$_5$ 
concentration indices separate well the unresolved foreground stars from extended
GCs and the majority of the contaminating background galaxies during the automated
GC selection. 

We have detected 57 old GC candidates in 17 dIrrs, 6 GCs in 2 dSph, 25
GCs in 2 dEs and 25 GCs in 4 Sm dwarfs. Therefore, in combination with
the 60 GCs from \cite{Georgiev08} we have in total 117 GCs in 30 dIrrs.
The total number of GCs in our sample contains: 119 in dIrrs, of which 
64 are bGCs and 42 rGCs; in dEs/dSphs the number of GCs sums up to 31 
with 21 bGCs and 10 rGCs; the low-mass late-type spirals contain 26 GCs 
in total, with 13 bGCs and 11 rGCs. The properties of all GC candidates 
are listed in Table\,\ref{table:GCCs} with the cluster ID and its 
coordiantes in columns (2) and (3); cluster absolute magnitude and 
foreground dereddened color $(V-I)_0$ in (3) and (4); cluster half-light 
radius and ellipticity ($r_{\rm h},\ \epsilon$, respectively) in (5) and 
(6) as derived in Sect.\,\ref{sizes} and their projected distance from 
the galactic center ($r_{\rm proj}$) and the normalized to the effective 
galaxy radius projected distance ($r_{\rm proj}/r_{\rm eff}$) in (7) and 
(8), respectively. Finding charts of all GCs are presented 
in the Appendix in Figure~\ref{color1}.

The final GC photometry was derived from curve of growth
analysis for each individual object from images iteratively cleaned from
contaminating sources within the photometric apertures. A detailed
description of this iterative procedure is provided in \cite{Georgiev08}
and will not be repeated here. 

Conversion from STMAG to Johnson/Cousins was performed using the
transformation and dereddening coefficients (for a G2 star) provided in
\cite{Sirianni05}. The Galactic foreground reddening $E_{B-V}$ toward 
each galaxy was obtained from the \cite{Schlegel98} maps.

\subsection{Structural parameters}\label{sizes}
To derive the GC structural parameters (half-light radius, $r_{\rm h}$,
and ellipticities, $\epsilon$) we have used the {\sc ishape} algorithm
\cite[]{Larsen99}, which models the object's surface brightness profile
with analytical models convolved with a subsampled point spread function (PSF). Since most
clusters are marginally resolved, a very good knowledge of the ACS PSF is
required to obtain reliable measurements of their sizes. Testing the
influence of the PSF library, we performed a comparison between the
$r_{\rm h}$ measured using {\sc
TinyTim}\footnote{http://www.stsci.edu/software/tinytim/tinytim.html \\
The {\sc TinyTim} software package takes into account the field-dependent
WFC aberration, filter passband effects, charge diffusion variations, and
varying pixel area due to the significant field distortion in the ACS
field of view \cite[]{Krist&Hook04}.} model PSFs (TT-PSF) and the
empirical, local PSF (lPSF), built from stars in the image of each galaxy.

The ten times sub-sampled TT-PSF was convolved with the charge diffusion
kernel. The position-dependent lPSFs was built from images typically
having more than 20 isolated stars across the ACS field. For a few galaxies
with a very low point-source density, we use the lPSF library from
ESO~223-09, which has the maximum number of good PSF stars of all galaxies
in our sample (69). Both PSF types were used with {\sc ishape} to model the
cluster profiles for the all available King model concentration parameters
($C=r_{\rm tidal}/r_{\rm core}=5,15,30,100$). The final cluster parameters were
adopted from the best $\chi^2$ fit model for both PSFs. For the final
cluster half-light radius we have adopted the geometric mean from the {\sc
ishape} FWHMs measured along the semi-major and semi-minor axis ($w_{y}$
and $w_{y}$), i.e. $r_{\rm h}=r_{{\rm h}, w_{y}}\sqrt[]{w_{x}/w_{y}}$.

\begin{figure}
\begin{center}
\epsfig{file=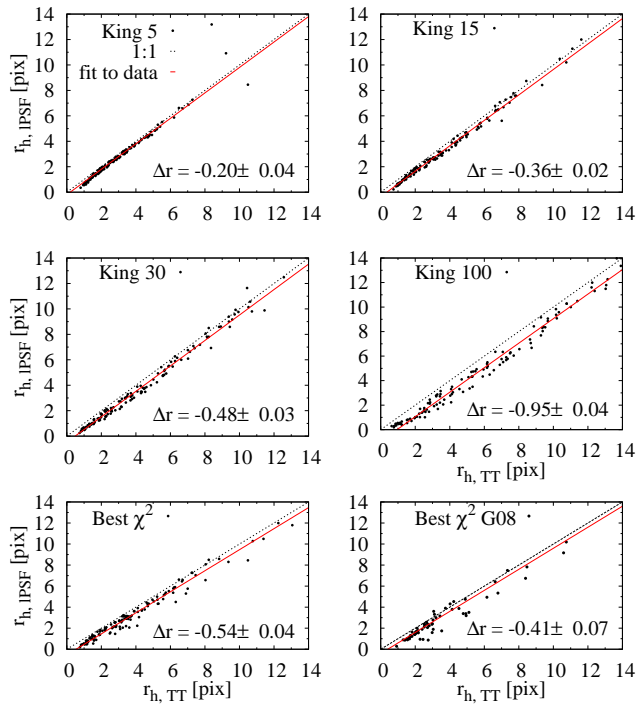, width=0.5\textwidth, bb=50 50 600 650}
\caption{Comparison between $r_{\rm h}$ derived with {\sc TinyTim} (TT)
and empirical PSFs build from the images (lPSF). From top to bottom are
shown relations for different King model concentration parameters and the
best $\chi^2$ fit models, respectively. The lower right panel shows the
best $\chi^2$ from our previous study \protect\cite[]{Georgiev08}. 
The solid line shows the 1:1 relation.} \label{TTvslPSF}
\end{center}
\end{figure}

Figure\,\ref{TTvslPSF} shows a comparison between the $r_{\rm h}$ derived
from the best-fit TT-PSF and lPSF models. It is obvious that for
every concentration parameter  the $r_{\rm h, lPSF}$ values are smaller
than the $r_{\rm h, TT}$ values (see $\Delta r$ values in
Fig.\,\ref{TTvslPSF}). This shows that the {\sc TinyTim} PSFs are sharper
than the empirical lPSFs which represent the actual imaging conditions (incl.~charge
diffusion, telescope focus breathing, drizzle effects, etc.).

In this study we are compiling data from our previous study \cite[]{Georgiev08} in
which the $r_{\rm h}$ were derived from the best $\chi^2$ TT-PSF due to the lack
of enough good PSF stars for modeling the lPSF in the field of those extremely
isolated dIrrs. Therefore, we convolved the GCs from that study with the local PSF
derived from the image of ESO\,223-09. We point out that variable telescope focus
changes may introduce unknown systematics. In the bottom right panel of
Fig.\,\ref{TTvslPSF} we show the $r_{\rm h}$ derived with the {\sc TinyTim} PSF
versus the empirical PSF for the best $\chi^2$ model. As can be seen, the values
computed with TT-PSFs are $\sim0.4$\,pix larger than the corresponding sizes
computed with the empirical lPSFs. 

A similar difference between the empirical lPSF and the TT-PSF of
$\sim0.5$\,pix (the $r_{\rm h, TT}$ being larger) was also previously
reported in the ACS study of the Sombrero galaxy by \cite{Spitler06}
(their Fig.2), where a fixed concentration index $C=30$ was assumed. In
the following, we will use the structural parameters of GCs obtained with
the empirical lPSFs.

\subsection{Completeness Tests}\label{comptest}

\begin{figure}
\begin{center}
\epsfig{file=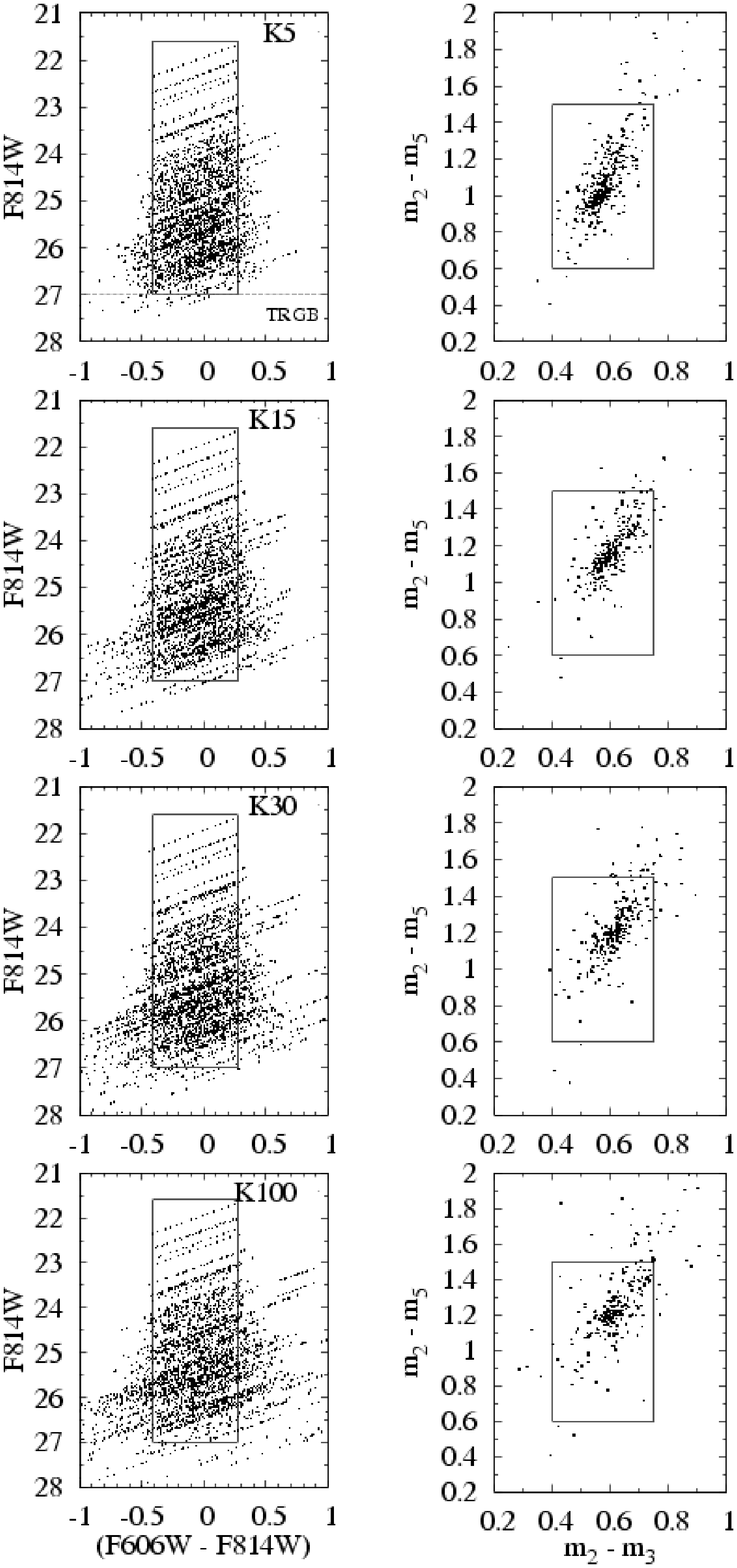, width=0.6\textwidth, bb=844 134 1430 1100}
\caption{{\it Left panels:} Color-magnitude diagrams of the artificial clusters with 
King profile concentrations $C=5, 15, 30$ and 100 from top to bottom, 
respectively. The solid line rectangle indicates the color-magnitude region 
used to generate artificial clusters and their subsequent selection. {\it
Right panels:} Magnitude concentration index of the artificial clusters as defined by
their 2,3 and 5\,pixel aperture radius magnitudes. The rectangle defines the 
final cluster selection.}\label{select}
\end{center}
\end{figure}

\begin{figure}
\begin{center}
\epsfig{file=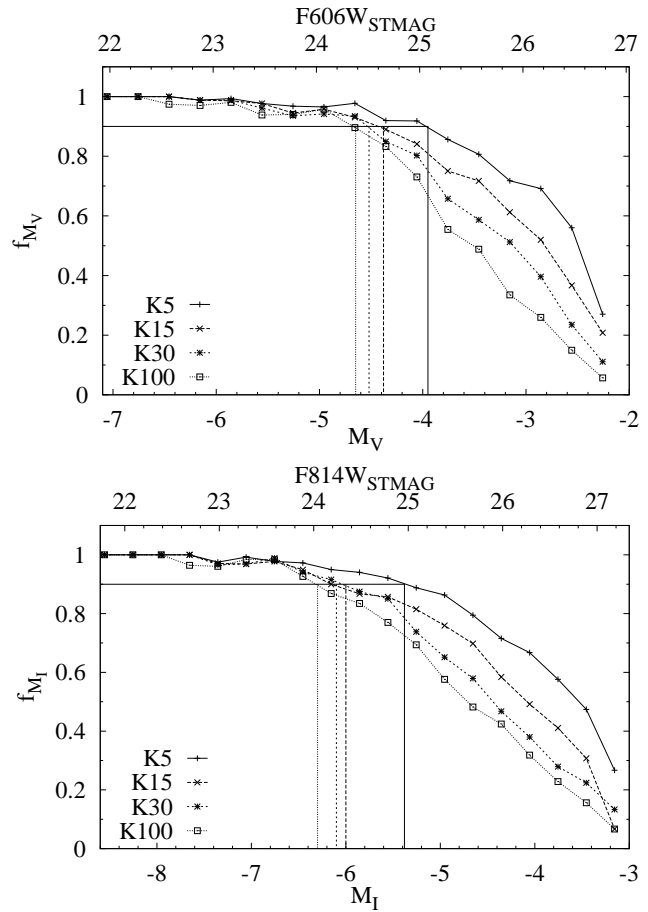, width=0.6\textwidth, bb=60 70 490 550}
\caption{Completeness as a function of synthetic cluster magnitude for $V$ and $I$
in the top and bottom panels, respectively. To convert from instrumental STMAG
to the Johnson/Cousins photometric system, we have used the transformation 
relations tabulated in \protect\cite{Sirianni05}. To convert from apparent to absolute
magnitudes, we have used the distance modulus of $m-M=29.06$\,mag to 
ESO\,223-09 as derived from the TRGB measurement by 
\protect\cite{Karachentsev07}.
\label{completeness}
}
\end{center}
\end{figure}

Artificial clusters were modeled with the {\sc mkcmppsf} and {\sc mksynth}
routines in the {\sc baolab} package \cite[]{Larsen99}. The procedure
includes convolution of empirical PSFs with analytic KING profiles for all
concentrations ($C=5$, 15, 30 and 100). For the completeness tests we
chose the galaxy ESO\,223-09 which has the highest surface brightness,
foreground extinction ($E_{B-V}\!=\!0.260$) and foreground star contamination
in our sample (cf. Fig.~\ref{color1}). We point out that this is the most
conservative estimate of the photometric completeness among our sample galaxies,
i.e., all other sample galaxies have more complete GC samples. To sample the
parameter space typical for colors and luminosities of old GCs ($t\ge4$ Gyr) we generated
synthetic clusters in the range $21<$\,F606W\,$<26$\,mag (STMAG) with the
fainter limit matching the TRGB at the distance of ESO\,223-09
\cite[]{Karachentsev07}. The colors of the artificial clusters were spread
over the range $-0.4<F606W-F814W<0.25$ ($0.5<V-I<1.5$) in fourteen color
bins with 0.05\,mag step. For each color bin 200 arificial clusters were
randomly placed across the synthetic image. This step was repeated 100
times thus providing 20\,000 objects per color bin or in total 280\,000
artificial clusters for the completeness test for one King profile
concentration. The same procedure was applied to all four concentration
parameters provided within {\sc baolab}.

The synthetic images consist of the modeled clusters on a constant (0
ADUs) background, matching the mean background value of the original ACS
images. Every synthetic image, containing 200 objects, was then added to
the original ESO\,223-09 image, thus preserving the image background level
and noise characteristics. The final images were then subjected through
the same routines for object detection, photometry, and GC selection as
the ones used to define the science sample.

In Figure~\ref{select} we show the color-magnitude diagrams (CMDs) and
concentration parameter plots of the retrieved artificial clusters for the four King
models. The rectangle in the left-column panels indicates the region in
which the artificial clusters were modeled and later selected. After
applying the color-magnitude and the {\sc imexam} FWHM and $\epsilon$
cuts, the final cluster selection was based on their concentration parameters as
derived from the difference between their magnitudes in 2, 3 and 5\,pixel
aperture radius. The right-column panels of Figure\,\ref{select} show the
${\rm m}_2-{\rm m}_3$ vs. ${\rm m}_2-{\rm m}_5$ diagrams. The objects that
survived all the selection criteria were used to compute the completeness
as a function of the cluster magnitude. In Figure\,\ref{completeness} we
present the completeness functions in $V$ and $I$ for all King models.

Figure~\ref{completeness} shows that for the case of a galaxy with high surface
brightness, strong foreground reddening and foreground star contamination
(cf. Fig.~\ref{color1}), the $90\%$ completness limit for extended sources is
reached at $M_V\simeq-4.5$ mag. As expected, the completeness is a function of
the cluster size, in the sense that more extended clusters suffer
stronger incompleteness as their detection and correct magnitude measurement are
easily affected by the variable galaxy background and bright foreground stars. We
point out that this is the least complete case in our sample and that all other
target galaxies have fainter completeness limits. For all our target galaxies we
sample more than 99\% of the total GCLF, in terms of luminosity and mass.

\subsection{Background Contamination}\label{contamination}
Background contamination from bulges of compact galaxies 
at intermediate redshifts, which resemble the colors and structural
appearance of GCs need to be taken into account. We have already assessed
the expected contamination for the ACS field of view
and objects with identical magnitude and size distributions as GCs in
galaxies within 8\,Mpc using the Hubble Ultra Deep Field
(HUDF)\footnote{http://heasarc.nasa.gov/W3Browse/all/hubbleudf.html} \cite[]{Georgiev08}. The
expected number of contaminating background galaxies increases
significantly for $V_0\geq25$\,mag ($M_V\geq-4$), well beyond the
luminosity distribution of GCs. Nevertheless, down to this limit a
contamination of up to 2 objects per field is expected.

\section{Analysis}

\subsection{Integrated Galaxy Magnitudes}\label{SB}
Only six of the dwarf galaxies in our sample have total $V-$band magnitudes
available in the literature while the majority of them only have $B$
magnitudes. However, good knowledge of the $V$-band magnitudes is required to
estimate their GC specific frequencies for a robust comparison with previous
studies. Hence, we performed integrated-light photometry on the ACS images and
derived their total magnitudes. 

Bright foreground stars and background galaxies were masked out and
replaced with the local background level and noise. The median smoothed
image of each galaxy was used to estimate the center of the galaxy.
However, for most dwarfs the derived centers were not representative of
the visual center of the extended galaxy light, but rather the region with
the strongest starburst. Thus, we adopted the geometric center of the
isointensity contour at the $10\sigma$ level above the background as the
galaxy center. Since the galaxies rarely extend beyond 1500\,pixel radius,
the estimate of the sky value determined from the median value measured at
the image corners, is representative for the true photometric background.
To measure the galaxy magnitudes we used the {\sc ellipse} task within
IRAF. The initial values for galaxy ellipticity and position angle were
taken from NED. For dwarfs without published values for those parameters,
we estimated the center in interactive mode with {\sc ellipse}, i.e. to
approximate the ellipticity and position angle of the extended light (at
the $10\sigma$ isointensity contour). We have measured the total galaxy
magnitudes within the ellipse with radius at the Holmberg radius
($\mu_{B}=26.5$) quoted in NED.

Magnitudes were adopted from NED for two spiral galaxies (ESO\,274-01 and
NGC\,247, see Table\,\ref{glist}) which were extending beyond the ACS field of
view as well as for NGC\,4605 which was off-center and for KKH\,77 which was
contaminated by a very bright (and saturated) foreground star. These magnitudes
from NED are (deprojected) total magnitudes corrected for internal reddening
assuming $E_{B-V}=0.05$ mag. For IKN and VKN, two extremely low surface brightness
dwarfs ($M_V\approx-11.5$ and $-10.5$\,mag, respectively), we could not reliably
determine their magnitudes due to an extremely bright foreground star in the
former and the very low surface brightness of the latter (close to the level of
background fluctuations and measurement errors). Their $V-$band magnitudes were
derived from their $B-$band magnitudes assuming an average $B-V=0.45$ mag
estimated from the rest of the dwarfs. Due to the fact that UGCA\,86 was centered
in the middle of ACS chip 2, its Holmberg radius extended beyond the ACS field by
$\sim\!1\arcmin$ and, therefore, we had to extrapolate its magnitude with the
gradient of the last 3 magnitude bins of its curve of growth. 

Finally, the derived magnitudes were compared with the published ones for
six dwarfs in our sample (DDO\,52, ESO\,269-58, IC\,4662, NGC\,5237, 4068,
4163) and found consistent within the measurement errors. The magnitudes
of all galaxies are listed in Table\,\ref{glist}.

\begin{figure*}
\epsfig{file=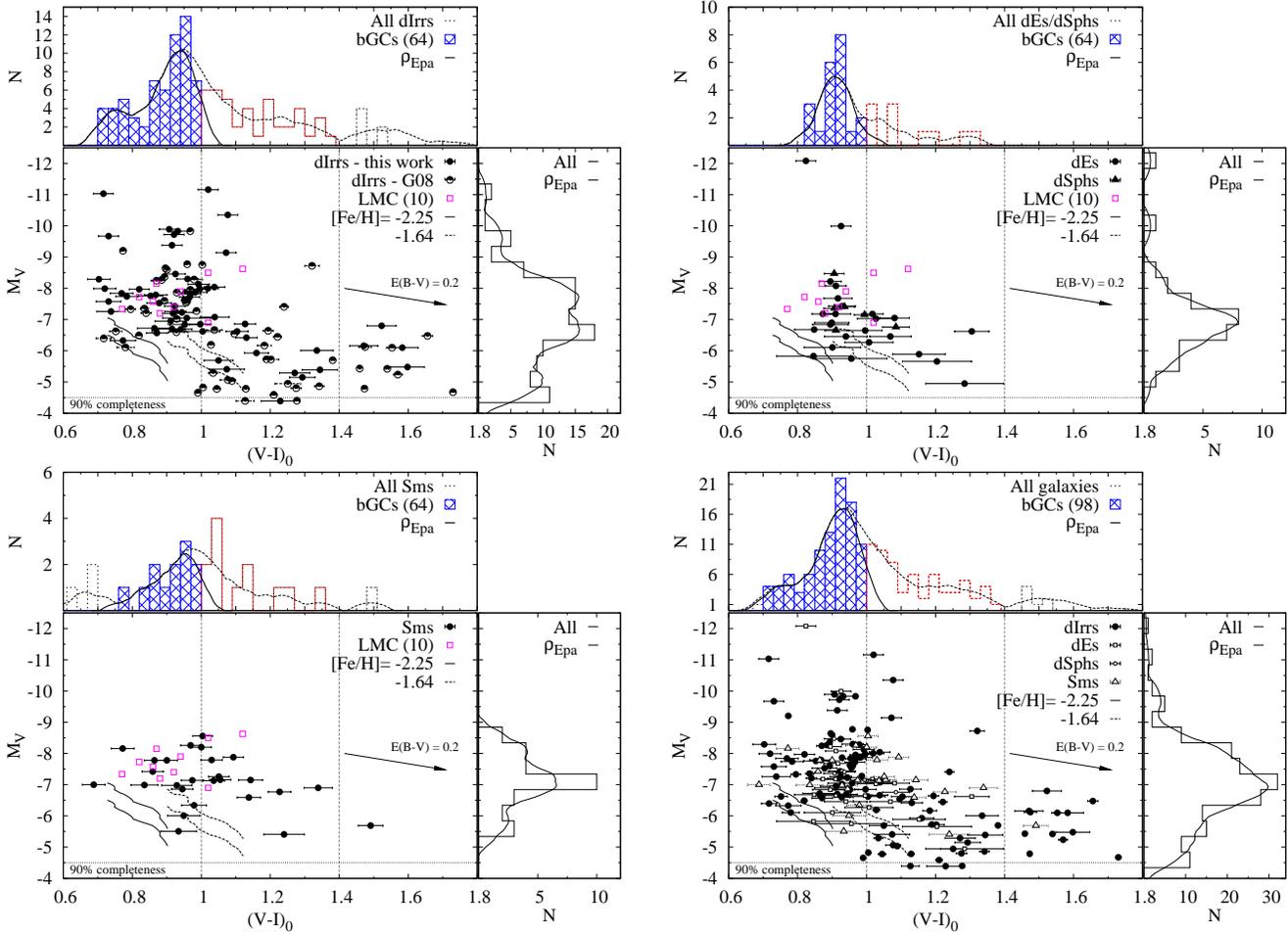,width=1\textwidth, bb=50 50 968 717}
\caption{Combined color-magnitude diagrams for GC candidates in our sample
galaxies split by the morphological type of the host. GCs in dIrr
galaxies are shown in the upper left panel; the other panels show the
corresponding distributions for dEs/dSphs, Sm, and of the combined
sample. Filled and semi-filled circles in the dIrr-panel indicate GC
candidates from the current and \protect\cite{Georgiev08} study. Open
squares are old LMC GCs \protect\cite[data from][]{McLaughlin&vdMarel05}.
The vertical dotted lines mark the color separating blue and red GC
candidates at $V\!-\!I=1.0$ mag. Thick solid and dashed curves show
\protect\cite{BC03} SSP evolutionary tracks for two clusters masses
$M_{\rm cl}=3\cdot10^4$ and $5\cdot10^4\,M_\odot$ from 3 to 14 Gyr (left to right)
and metallicities [Fe/H]~$=-2.25$ and $-1.64$ dex, respectively.
Histograms illustrate the GC candidate color and luminosity distributions. A reddening vector shows the effect of $E_{B-V}=0.2$ of internal extinction.
\label{CMD}
}
\end{figure*}

\subsection{Color Distributions}\label{CMDsect}
In Figure~\ref{CMD} we show the color-magnitude diagrams for GCs in dIrr,
dE/dSph, and Sm galaxies, and the composite sample, combining the results
from this work with our identically analyzed sample from
\cite{Georgiev08}. The color and luminosity distributions are illustrated
as histograms in the top and right sub-panels, where the curves indicate
non-parametric Epanechnikov kernel probability density estimates. We
subdivide the samples in color into blue GCs (bGCs) with
$V\!-\!I\leq1.0$ mag and red GCs (rGCs) with $V\!-\!I>1.0$ mag, which
include the sub-sample of extremely red objects with colors $V\!-\!I>1.4$
mag\footnote{The faintest clusters with $M_V\leq-6$ and
$(V\!-\!I)_0>1.0$\,mag are likely background contaminants which passed our
selection criteria. Their nature will be confirmed by follow-up spectroscopy.}.
This division is motivated by the average location of the gap between the
blue and red color peaks of rich GCSs in massive early-type galaxies
\citep[e.g.][]{Gebhardt&Kissler-Patig99}.

An intriguing feature of Figure~\ref{CMD} is the lack of faint ($M_V \ga -6$) blue
GCs in our sample dIrr galaxies. Our artificial cluster tests show that this
is not a completeness effect as our 90\% completeness limit is at
$M_V=-4.5$\,mag (see Sect.~\ref{comptest}). Finding one or more GCs in
these faint dwarfs would increase their already high specific frequencies
even more (the specific frequencies will be discussed in a subsequent paper of this series). On the other hand, we do observe
few clusters at $M_V\simeq-5.5$ to $-6.5$ mag and $(V\!-\!I)<1.0$ mag in
galaxies with dE/dSph and Sm morphology\footnote{We note that all absolute
magnitudes were calculated using the newly determined distance moduli by
\cite{Tully06} and \cite{Karachentsev06, Karachentsev07} which are fainter
by $\sim\!0.5$\,mag on average from the values listed in
\cite{Karachentsev04}, hence, the GCs have brighter absolute
magnitudes.}.

One plausible explanation for the apparent lack of faint blue clusters 
is that the metal-poor ($V-I<1.0$ mag) and low-mass GCs ($M_V>-6.5$ mag) 
are actually younger (age $<4$ Gyr) than our selection criteria. If this 
is spectroscopically confirmed would imply that GCs in dIrrs formed a bit 
later than GCs in more massive early-type galaxies. To explore this effect 
we investigate stellar evolution fading according to the SSP model
of \citet{BC03}. Passive aging of a simple stellar population from 3
to 14\,Gyr reddens its $(V-I)_{0}$ color by $\sim0.15$\,mag and fades its
$V-$band luminosity by $\sim1.5$\,mag. This is illustrated in
Figure\,\ref{CMD} with evolutionary tracks for two metallicities
([Fe/H]~$=-2.25$ and $-1.64$ dex) and two cluster masses ($M_{\rm
cl}=3\cdot10^4$ and $5\cdot10^4\,M_\odot$)
If some of the bluest GCs in our sample are indeed younger
clusters (with $t\la 4$ Gyr) they will end up on the faint-end of the
luminosity function at an age of 14 Gyr. 
However, those clusters would have to have unusually low metallicities
([Fe/H]~$\le-2.0$), an interesting result that calls for spectroscopic
confirmation. Previous spectroscopic analyses of GCs in other dwarf
galaxies show that their blue colors are in general consistent with old
ages and low metallicities. However, some of these clusters show
spectroscopic intermediate ages ($\sim4$\,Gyr), in particular in dIrrs
\cite[]{Puzia&Sharina08}. An alternative (though perhaps less likely) explanation
for the lack of old, metal-poor, low-mass clusters in dIrr galaxies may be
selective reddening of such objects. The reddening vector in the CMDs of
Figure~\ref{CMD} 
shows that a reddening of $E_{B-V}\!=\!0.2$ mag is enough
to dislocate intrinsically blue GCs to the red GC sub-sample. The age and
reddening effects should be tested with spectroscopic observations.

The probability density estimates, as shown in Figure~\ref{CMD}, for all galaxy 
subsamples give the highest probability values in the range
$(V-I)_{0,\rho}\approx0.9-1.0$\,mag and $M_{V,\rho}\approx-7.5$ to $-6.5$
mag. Gaussian fits to the smoothed bGC luminosity function give peak values in 
the range $M_{V,{\rm TO}}=-7.6$ to $-7.0$ mag with $\sigma_{\rm GCLF}=1.2 - 1.5$
mag (see Sect.\,\ref{ln:lumfunc}). The color distributions of GCs in our sample 
galaxies peak at values typically found in other low-mass dwarfs \cite[][]{Seth04, 
Sharina05, Georgiev06, Georgiev08}, and are very similar to the canonical blue peak 
color of rich GCSs in massive early-type galaxies \citep[e.g.][]{Peng08}. For 
comparison we also show ten of the brightest old LMC GCs \citep[]{McLaughlin&vdMarel05}. 
There are three more LMC clusters that are fainter than $M_V=-6.5$ mag, however, 
without available $V\!-\!I$ colors. 

\begin{figure}
\epsfig{file=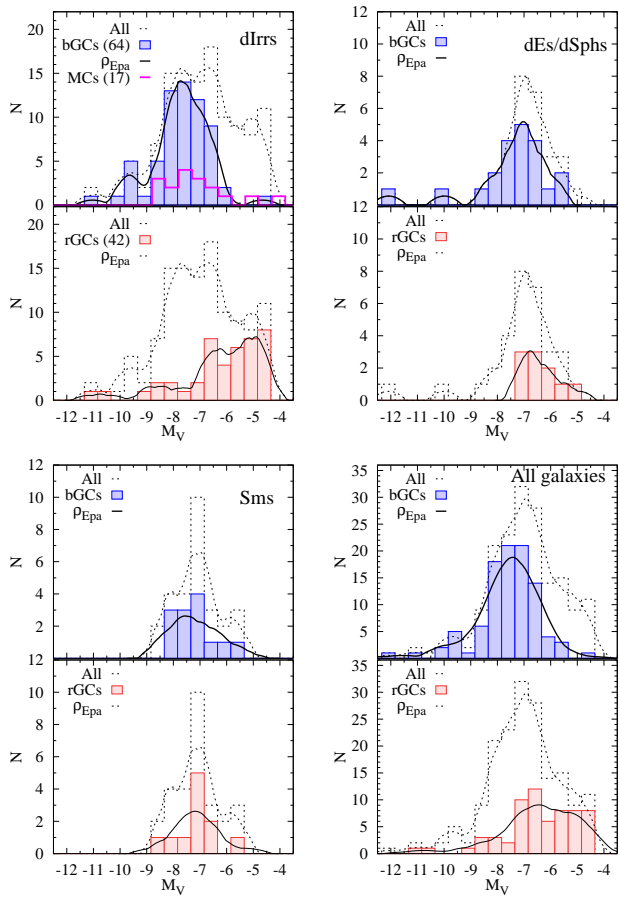,width=0.5\textwidth, bb=50 60 420 554}
\caption{Luminosity distributions of selected blue and red GCs in the
top and bottom panels, respectively. Each sub-sample was split at 
$V-I\!=\!1.0$ mag in bGCs and rGCs. Shaded histograms show the
corresponding GC candidates luminosity distribution, while open histograms
are the total samples for a given host morphology. In all panels, thick
and dotted curves are non-parametric Epanechnikov-kernel probability
density estimates. The solid line open histogram shows in the upper left 
panel the luminosity distribution of old LMC GCs for comparison 
\protect\cite[data from][]{McLaughlin&vdMarel05}.
\label{MVI_hist}}
\end{figure}

\subsection{Luminosity Functions}\label{ln:lumfunc}
Figure~\ref{MVI_hist} shows the luminosity functions of the blue and red GCs in
the top and bottom sub-panels, respectively, which were split at $V-I\!=\!1.0$
mag. Thick curves are Epanechnikov kernel probability density estimates. We find 
that the rGCs are biased toward fainter luminosities compared to the bGC sub-sample. 
This indicates that these objects are either strongly affected by background 
contamination or intrinsically fainter than their bGC counterparts. Gaussian 
fits to the smoothed histogram distributions return $M_{V, {\rm TO}}=-7.56\pm0.02$\,mag 
and $\sigma_{\rm GCLF}=1.23\pm0.03$ for dIrr, $M_{V, {\rm TO}}=-7.04\pm0.02$\,mag 
and $\sigma_{\rm GCLF}=1.15\pm0.02$ for dE/dSph, and $M_{V, {\rm TO}}=-7.30\pm0.01$\,mag 
and $\sigma_{\rm GCLF}=1.46\pm0.02$ for Sm galaxies (note the different sample sizes 
when comparing sub-populations in Fig.\,\ref{MVI_hist}). Assuming a typical 
$M/L_V=1.8$ obtained for old metal-poor Magellanic GCs \cite[]{McLaughlin&vdMarel05} 
the turnover magnitude translates to a turnover mass $m_{\rm TO}\simeq1.6\times10^5$\,M$_\odot$, in excellent
agreement with the results of \cite{Jordan07}. 

The GC luminosity function turn over magnitude for dIrrs is slightly brighter than
those for dE/dSph and Sm galaxies, and it shows significantly broader luminosity
function peaks extending to fainter magnitudes. This may be due to the interplay
of different formation mechanisms and ages/metallicities or perhaps due to
contamination by background galaxies. In general, all $M_{V, {\rm TO}}$ values are
consistent with the luminosity function turn-over magnitude for metal-poor
Galactic GCs \citep{DiCriscienzo06}, as well as for GCs in early-type dwarfs
\citep{Sharina05,Jordan07,Miller&Lotz07}, and virtually identical to the turn-over
magnitude of old LMC GCs at $M_V=-7.50\pm0.16$\,mag \cite[data
from][]{McLaughlin&vdMarel05}. 

\subsection{Nucleated Dwarf Irregular Galaxies}
Another interesting feature in the CMDs of Figure~\ref{CMD} is the presence of a
few relatively bright GCs in dIrr and dE/dSph galaxies, which are similar in color
and magnitude to $\omega$\,Cen and M\,54. Those clusters are located in the nuclear regions of
their host galaxies. Such bright objects do not appear in the Sm sub-sample. A
dedicated study of the properties of these bright GCs will be presented in a
forthcoming paper of this series. 

\subsection{Structural Parameters}\label{structpar}
Since the clusters' half-light radii, $r_{\rm h}$, and ellipticities, $\epsilon$,
are stable over many relaxation times \cite[e.g.][]{Spitzer&Thuam72} and, thus,
contain information about the initial conditions and the dynamical evolution of
clusters over a Hubble time. In particular, the cluster half-light radius is
stable during $>\!10$ relaxation times (i.e., $\sim$\,10 Gyr)
\cite[e.g.][]{Aarseth&Heggie98}, while the ellipticity decreases by a factor of
two within five relaxation times and reaches asymptotic values around 0.1
\cite[]{Fall&Frenk85,Han&Ryden94,Meylan&Heggie97}. 

\begin{figure*}
\epsfig{file=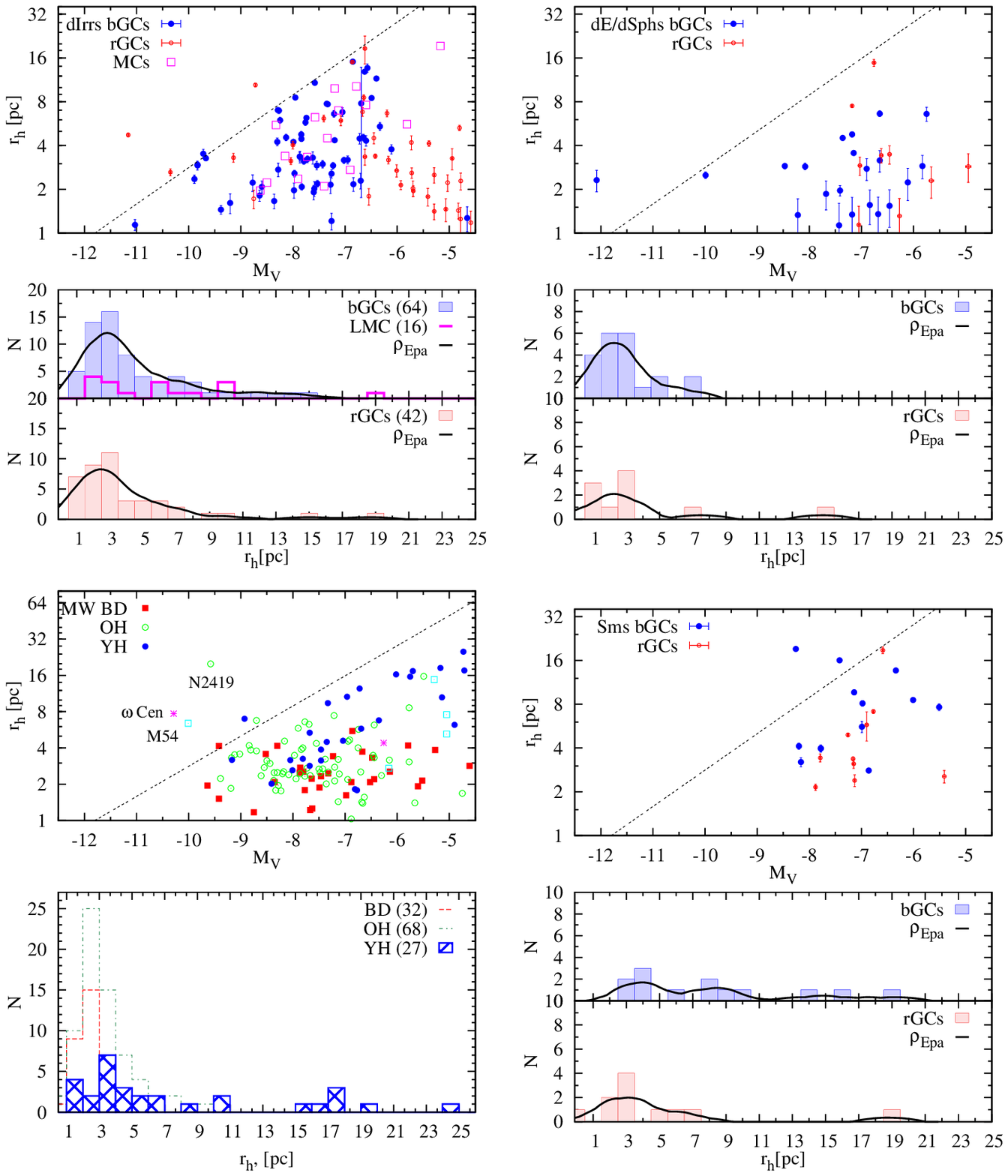, width=14cm, bb=50 50 470 554}
\caption{Luminosity $M_V$ versus half-light radius $r_{\rm h}$ for GCs in dIrrs {\it upper 
left panel}, dE/dSph {\it upper right panel}, and Sm galaxies {\it lower right 
panel}. Each sub-sample was split at $V-I\!=\!1.0$ mag in bGCs and rGCs. In 
the upper left panel we show old GCs in the Magellanic Clouds \citep{McLaughlin&vdMarel05}. 
The dashed line indicates the upper envelope for the distribution of Galactic 
GCs \citep{Mackey&vdBergh05}. The lower sub-panels show the corresponding 
$r_{\rm h}$ distributions for the blue and red GC candidates. The thick curves 
are Epanechnikov-kernel probability density estimates. Note that in all our 
sub-samples the $r_{\rm h}$ distributions of blue GCs appears more extended 
than that of the red GCs.
\label{rh}}
\end{figure*}

\subsubsection{Half-Light Radii}
In Figure~\ref{rh} we present the measurements of $r_{\rm h}$ for GCs in our
sample galaxies. The majority of GCs lies below the empirically established
relation $\log r_{\rm h}=0.25\times M_V + 2.95$ \citep{Mackey&vdBergh05}, 
which forms the upper envelope of Galactic GCs in the $M_V$\,vs.\,$r_{\rm h}$
plane (see lower left sub panels in Fig.\,\ref{rh}). Some of these brightest 
GCs, that reside in the nuclear regions of their hosts, tend to lie on or 
above that envelope (towards larger $r_{h}$ at a given $M_{V}$), which seems 
typical for nuclear star clusters \cite[e.g.][]{Boker04,Hasegan05}. This 
region is also occupied by the peculiar Galactic GCs $\omega$\,Cen, NGC\,2419, 
NGC\,2808, NGC\,6441 and M\,54, the nucleus of the Sagittarius dSph galaxy. 
For an object with high S/N ($>50$) and good knowledge of the PSF, the 
theoretical spatial resolution limit can be as good as $10\%$ of the PSF 
($\sim0.2$\,pix for ACS/PSF) \cite[]{Larsen99}. Thus, at the median distance 
of the entire galaxy sample of $\sim5$\,Mpc and taking into account the $r_{\rm h}$ 
measurement error, the spatial resolution can be as good as $\sim0.9$\,pc 
($\sim0.8$\,pix). Therefore, the majority of the clusters with high S/N are 
well resolved, even for the distant most galaxy in our sample at $\sim12$\,Mpc.

The half-light radius distribution of the bGCs and rGCs is shown in the bottom
sub-panels of Figure\,\ref{rh}. On average, bGCs appear more extended than rGCs in
all sub-samples, by about 9\%, however with low statistical significance. On
average, the most compact bGC population is found in dE/dSph host galaxies
($r_{\rm h,med}=2.5$\,pc), followed by bGCs in dIrrs ($r_{\rm h,med}=3.3$\,pc) and
Sm galaxies  ($r_{\rm h,med}=7.6$\,pc), whose GC population is more incomplete due
to the restricted spatial coverage (cf. Sect.\,\ref{SB}). We find the highest
value of the probability density estimate of the entire sample at $r_{\rm
  h}\approx2.9$\,pc and a median $r_{\rm h,med}=3.2\pm0.5$\,pc. These values are
typical for Galactic GCs \citep{Harris01}. The median value of the old LMC GCs is
$r_{\rm h, med}=5$\,pc \citep[based on measurements
from][]{McLaughlin&vdMarel05}. 

A comparison between the $r_{\rm h}$ distribution of the total bGC
sample with the sizes of different metal-poor, Galactic GC sub-populations
shows that the old halo (OH) GCs have comparable ($r_{\rm h, med}=3$\,pc)
and the young halo (YH) GCs have larger median sizes ($r_{\rm h,
med}=5.4$\,pc) than blue GCs in our sample dwarf galaxies. We note, however, 
that if all Galactic GCs are considered, there is practically no difference 
in the mean $r_{\rm h}$ . In the light of the accretion origin of the YH-GCs, 
their $r_{\rm h}$ might have been influenced by the change of the host 
potential, which leads to an increase of the cluster $r_{\rm t}$ at large 
Galactocentric distances, and disk/bulge shocking causing mass loss, thus 
toward lower masses and larger $r_{\rm h}$ (cf Fig.\,\ref{rhMass}, 
Sect.\,\ref{Subsect.dynamical_evo}). This is supported by the high 
orbital energy (E$_{\rm tot}$, Z$_{\rm max}$, eccentricity, velocity, angular 
momentum) found for YH-GCs \cite[]{Lee07} which also supports an accretion 
origin of those.

\subsubsection{Ellipticities}
\begin{figure}
\centering
\epsfig{file=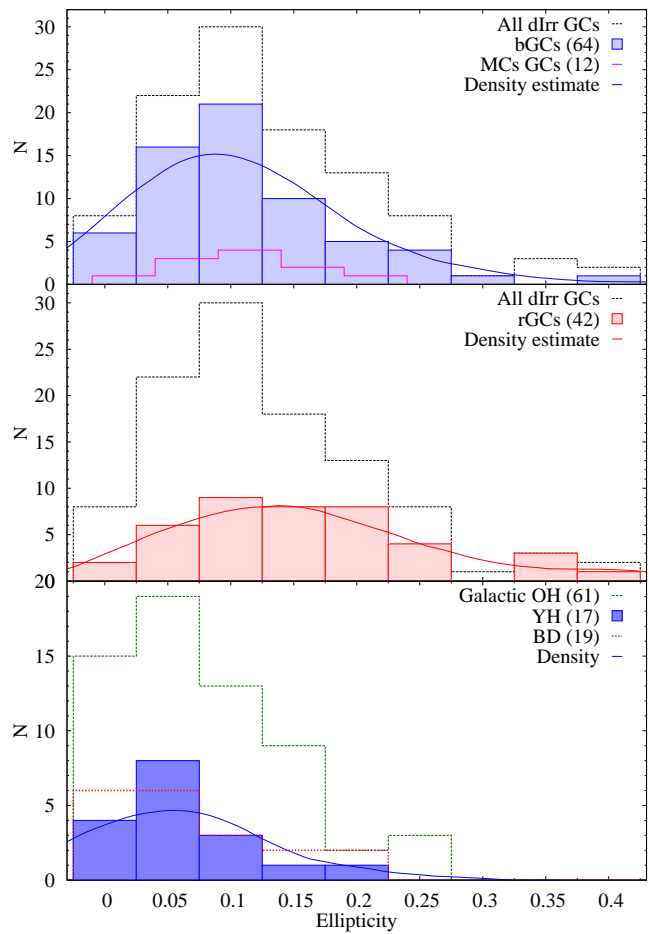, width=0.4\textwidth, bb=100 60 520 790}
\caption{Ellipticity ($\epsilon=1-b/a$) distributions of GCs in dwarf irregular galaxies. 
The top panel shows that bGCs in our sample dIrrs and in LMC are similarly flattened. 
The even broader $\epsilon$ distribution for the rGCs shows that the majority are 
likely background contaminants. The bottom panel shows the $\epsilon$ distribution 
for the Galactic GC subpopulations: OH = Old Halo; YH = Young Halo; BD =
Bulge/Disk. 
\label{ell_hist}}
\end{figure}

A difference between bGCs in dIrrs and those of the various Galactic GC
sub-populations is also found when clusters ellipticities are compared. In
Figure~\ref{ell_hist} we show the ellipticity distribution of GCs in our
sample dIrrs and in the Magellanic Clouds for which we used data from
\cite{Frenk&Fall82} and \cite{Kontizas89, Kontizas90}. The non-parametric
kernel density estimate identifies peaked distributions with values
$\epsilon\simeq0.1$ for both samples. This is a markedly different
ellipticity distributon than that of Galactic GCs which is biased towards
lower ellipticity values.

\cite{Fall&Frenk85} showed for self-gravitating clusters that during a period 
$\sim5t_{\rm rh}$ the ellipticity, $\epsilon$, decreases by a factor of two and
reaches an asymptotic value of $\epsilon\simeq0.1$, corresponding to the
observed mean ellipticity of our bGCs in dIrr galaxies. 
Under the assumption that the ages of the (bulk of the) GCs in our dIrr galaxies
are similar to those of Galactic GCs, 
this implies that our sample bGCs as well as the MC old GCs evolved dynamically in
isolation (i.e. mainly affected by internal processes rather than external cluster 
dissolution processes).

The broad ellipticity distribution of the rGCs, which extends toward 
large values, indicates that many or most of them are likely background 
contaminants.

\subsection{Dynamical State of Star Clusters in Dwarf Galaxies}
\label{Subsect.dynamical_evo}
The $r_{\rm h}$ versus cluster mass plane (Fig.~\ref{rhMass}) is often used to
illustrate cluster ``survivability'' that depends on the interplay between
various external and internal dissolution mechanisms
\cite[e.g.][]{Fall&Rees77, Gnedin&Ostriker97, Fall&Zhang01,
McLaughlin&Fall08}. In Figure\,\ref{rhMass} we show the distribution of
all sample GCs in this plane together with GCs in the Galaxy and the 
Magellanic Clouds (MCs). To convert from luminosities to masses we adopt 
a mean $M/L_V=1.8$ derived for the old MC clusters \citep{McLaughlin&vdMarel05}.
The half-light radius estimates and cluster masses for MC and Galactic
clusters were taken from \cite{McLaughlin&vdMarel05}, where available, 
and from \cite{Harris96} for the remainder. It should be noted, however, 
that if the GCs in our sample are on average younger, their $M/L-$ratios 
would be smaller.

\begin{figure*}
\begin{center}
\epsfig{file=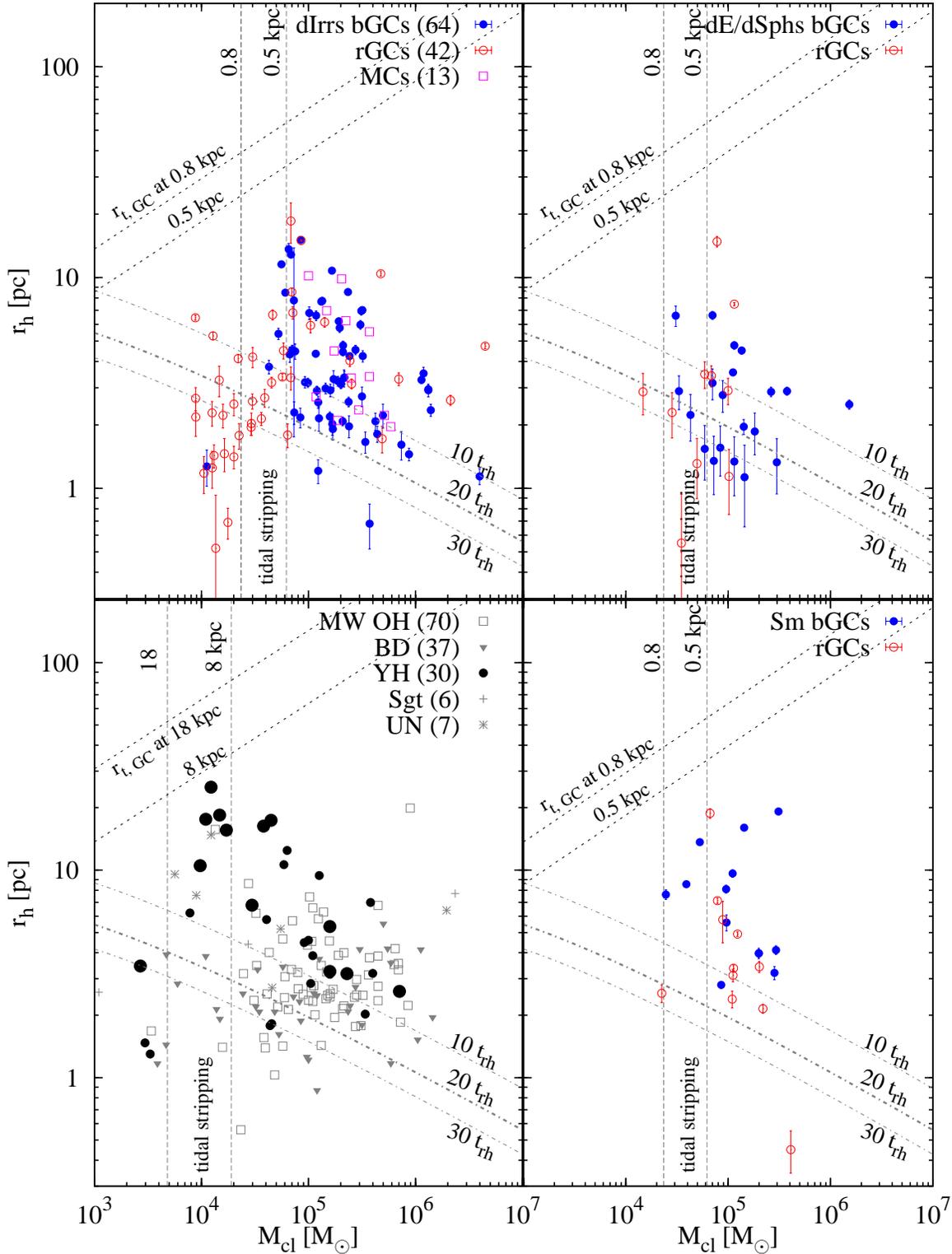, width=1.7\textwidth, bb=50 60 770 554}
\caption{Half-light radius versus custer mass for GCs in low-mass
galaxies. The panels show from the top left panel clockwise: bGCs and
rGCs (filled and open circles) in dIrr galaxies (split at $V-I=1.0$ mag)
and old Magellanic Clouds GCs (open squares); GCs in dE/dSph galaxies,
GCs in Sm galaxies; the lower left panel shows the Galactic YH-GCs, 
which were split according to their galactocentric distance being larger 
or smaller than 18\,kpc (large or small filled circles, respectively), 
and OH and BD GCs (open squares and filled triangles), Sagittarius dwarf 
GCs (plus signs), and clusters with unknown classification (asterisk). 
Lines indicate theoretical predictions for cluster dissolution over a 
Hubble time due to tidal field effects for clusters at 0.8 and 0.5\,kpc 
from the center of the host (left and right vertical lines, respectively), 
two-body relaxation processes as well as cluster re-expansion after 
(dash-dotted lines), (for details see Sect.~\ref{Subsect.dynamical_evo}). 
With dotted lines are also shown the cluster tidal radius $(r_{t,GC})$ at 
two assumed galactocentric distances (0.5 and 0.8\,kpc). Note that the
larger error bars in the dE/dSph panel are GCs in one of the most distant
galaxies in our sample - the dE UGC\,7369 at 11.6\,Mpc.
}\label{rhMass}
\end{center}
\end{figure*}

In the following we discuss the influence of the various dynamical effects
on the evolution of our sample GCs. Dash-dotted lines show the cluster
evaporation limit due to two-body relaxation \citep{Fall&Rees77,
Spitzer87, Gnedin&Ostriker97}
\begin{equation}
r_{\rm h}=\left(\frac{\rm 12\,000 {\rm [Myr]}}{20}\right)^{2/3}
\left(\frac{0.138}{\sqrt{G}m_\star {\rm ln}
\left(\gamma\frac{M}{m_\star}\right)}\right)^{-2/3}M^{-1/3}
\label{tevol}
\end{equation}
for clusters that survived 12\,Gyr after 20 initial relaxation times;
$m_\star=0.35 M_\odot$ is the average stellar mass in a GC after 12 Gyr of
evolution and $\gamma=0.02$ is a correction constant taken from models
which simulate clusters with a specific mass spectrum
\citep{Giersz&Heggie96}. GCs with $r_{\rm h}$ values lower 
values than this ``survival'' limit will have dissolved after a Hubble
time of dynamical evolution \citep[e.g.][]{Fall&Zhang01}.

The evolution of a star cluster in the Galactic tidal field was investigated by
\cite{Baumgardt&Makino03}. We have used their equation 7 to calculate the minimum
mass of a cluster which can survive for 12\,Gyr within a mean projected distance of $d_{\rm
  proj}\approx0.8$\,kpc from the galaxy center, matching the average $d_{\rm
  proj}$ of GCs in our sample. The result is shown as the left vertical dashed
line in Figure~\ref{rhMass} and corresponds to a minimum mass of
$\sim\!2.3\times10^4 M_\odot$ ($M_V=-5.5$\,mag). Clusters below this mass limit
would have dissolved in the tidal field of the dwarf galaxy after a Hubble time of
evolution. The right vertical line indicates the minimum cluster mass for $d_{\rm
  proj}=0.5$\,kpc. It should be mentioned, however, that effects due to a variable
tidal field and dynamical friction
were not taken into account. \cite{Vesperini00} showed that
low-mass galaxies efficiently disrupt the majority ($\sim90\%$) of their 
initial star-cluster population due to dynamical friction. 

An interesting aspect of cluster evolution is the re-expansion of star
clusters that can occur after the process of core collapse. The time
evolution of cluster mass and half-light radius is

\begin{equation}\label{eqCCexpan}
M(t) = M_{0}\left(\frac{t}{t_0}\right)^{-\nu}\ {\rm\ and}\ \ \ \ r_{\rm h}(t) = 
r_{\rm h_0}\left(\frac{t}{t_0}\right)^{\frac{2+\nu}{3}}
\end{equation}
and was first obtained by \cite{Henon65}, where $M=M_{cl}/m_{\star}$ and
$M_0$ and $r_{\rm h_0}$ are the cluster mass and half-light radius at the
time of core collapse $t_0$ and $\nu=0.01-0.1$ \cite[]{Baumgardt02} is the
dynamical mass-loss factor. Combining both equations one obtains

\begin{equation}\label{eq.rh}
r_{\rm h} = r_{\rm h0}\left(\frac{M}{M_0}\right)^{- \frac{2+\nu}{3\nu}}
\end{equation}
Since core collapse occurs within several initial cluster relaxation times
$t_0=n_{\rm rh}t_{\rm rh,i}$ \citep[see][]{Gieles&Baumgardt08} we can
combine Equations~\ref{tevol} and \ref{eq.rh} and obtain

\begin{equation}
r_{\rm h}=(t_0)^{2/3}\left(\frac{0.138}{\sqrt{G}m_\star 
{\rm ln}\left(\gamma\frac{M_0}{m_\star}\right)}\right)^{-2/3}M_{0}^{-1/3}
\left(\frac{M}{M_0}\right)^{- \frac{2+\nu}{3\nu}}
\end{equation}
Consequently, if $M_0\!=\!M$ (i.e.~no dynamical mass-loss after core
collapse) at $t_0=t=12$\,Gyr ($\sim$Hubble time), and $n_{\rm rh}=20$ the
equation yields the limit labeled ``$20 t_{\rm rh}$'' in
Figure~\ref{rhMass}. A cluster that starts its evolution below this
relation will be prone to expansion with a likely depletion of cluster 
stars prior to 20 relaxation times. In summary, the distribution of GCs 
in all studied dwarfs shows that their dynamical
evolution was governed by internal processes (two-body relaxation, stellar
evolution). The absence of GCs at masses lower than the lower mass limit
due to disruption by the galactic tidal field within the mean projected
radius of the GCs in the sample galaxies ($\sim\!3\!\times 10^4
M_{\odot}$, see Fig.~\ref{rhMass}) suggests that tidal disruption was
important as well. With dotted lines are in Figure\,\ref{rhMass} are 
shown the cluster tidal radius at two galactocentric distances, 0.5 and 
0.8\,kpc for the dwarfs in our sample and at 8 and 18\,kpc for Galactic 
clusters. This sets up an upper limit to the cluster size. The different 
distribution of rGCs with respect to that of bGCs in Fig. 8 indicates 
that most rGCs are background contaminants.

The comparison between our GC sample and Galactic GCs (bottom left panel
in Fig.~\ref{rhMass}) shows that bGCs, MC GCs, and Galactic YH GCs share
very similar distributions, which suggests that those objects have
experienced similar formation and/or dynamical evolution in the
weak tidal fields of the dwarf and the Milky Way halo regions. In
contrast, the Galactic OH and BD clusters show distributions that are significantly
different from those of dwarf galaxy bGCs and Galactic YH GCs. This may indicate
that their dynamical evolution was affected by the stronger inner Galactic tidal
field and disk/bulge shocking.


\section{Conclusions}
We present the analysis of archival F606W and F814W HST/ACS data for 68 low-mass
faint ($M_V\!>\!-16$ mag) dwarf galaxies located in the halo regions of nearby
($\la\!12$ Mpc), loose galaxy groups. Most of the dwarf galaxies in our sample are
more than 2\,mag fainter than the LMC ($M_V\!=\!-18.36$\,mag) and just as bright
as the SMC ($M_V\!=\!-16.82$ mag). The morphological makeup of our sample, summarized 
in Table\,\ref{tab:sum}, comprises 55 dIrrs, 3 dEs, 5 dSphs and 5 low-mass late-type 
dwarf spirals. Old GC candidates were found in 30 dIrrs,
2 dEs, 2 dSphs and in 4 Sms. In total we found 175 GC candidates and measure their
colors and magnitudes which are consistent with old and metal-poor stellar
populations. The total sample contains 97 blue GCs ($0.7<(V-I)_0<1.0$\,mag) GCs,
63 red GCs ($1.0<(V-I)_0<1.0$\,mag) with the rest being very red
($(V-I)_0>1.0$\,mag) likely background contaminants.

The combined color distribution of GCs in dIrrs peaks at 
$(V\!-\!I)\!=\!0.96\pm0.07$\,mag and the GC luminosity function turnover 
is at $M_{V,TO}\!=\!-7.6$\,mag, similar to the old LMC GCs. Gaussian fits 
to the smoothed histogram distributions return $M_{V,{\rm TO}}\!=\!-7.56\pm0.02$\,mag and
$\sigma\!=\!1.23\pm0.03$ for dIrrs, $M_{V,{\rm TO}}\!=\!-7.04\pm0.02$\,mag and
$\sigma\!=\!1.15\pm0.03$ for dE/dSph, and $M_{V,{\rm TO}}\!=\!-7.30\pm0.01$\,mag and
$\sigma\!=\!1.46\pm0.02$ for Sm galaxies. We thus find a tentative trend of
$M_{V,{\rm TO}}$ becoming fainter from late-type to early-type dwarf
galaxies.
Our artificial cluster tests show that this trend is not due to
incompleteness and may reflect relatively younger GC systems in dIrrs by
$\sim\!2-5$ Gyr depending on the metallicity. If confirmed, this would imply 
that GCs in dIrrs formed later than blue GCs in massive galaxies. Thus we 
suggest that this result be followed-up with spectroscopic observations.

The comparison of GC structural parameters with theoretical cluster disruption
models (including dynamical processes, such as stellar mass loss, relaxation,
tidal shocking indicates that the dynamical evolution of blue GCs in our sample is consistent with the evolution of
self-gravitating systems in a weak tidal field. Their half-light radii and cluster
masses evolve primarily due to two-body relaxation. The $r_{\rm h}$ vs. cluster
mass plane shows a similar distribution between bGCs in our sample and Galactic
Young Halo (YH) clusters, which is indicative of a similar formation and dynamical
evolution history. Dynamical models of star clusters evolving in isolation show
that they reach an asymptotic ellipticity $\epsilon\approx0.1$ after few cluster
relaxation times ($>5t_{\rm rh}$). Our analysis reveals that bGCs in our sample
galaxies on average more flattened ($\bar\epsilon=0.1$) than Galactic GCs, but
have a similar ellipticity distribution as GCs in the LMC. This suggests that old
GCs in low-mass galaxies are dynamically evolved stellar systems that spent most
of their evolution in benign tidal environment. 

We briefly report on the discovery of several bright (massive and/or young) GCs in
the nuclear regions of some dIrr and dSph galaxies. These massive nuclear clusters
show similar structural parameters as the peculiar Galactic clusters suspected of
being the remnant nuclei of accreted dwarf galaxies, such as M~54 and $\omega$Cen. 
If such accretion events happened early in the assembly history of the Galaxy, 
any tidal streams from the host low-surface brightness dIrr galaxies would likely 
have largely dissolved by now and thus escape detection. We will present a more 
detailed analysis of the properties of the nuclear clusters in a forthcoming paper 
in this series.


\section*{Acknowledgments}
IYG is grateful for the award of a STScI Graduate Research Fellowship, and
acknowledges support from the German Science Foundation through the grant
DFG-Projekt BO-779/32-1. THP acknowledges financial support from the
National Research Council of Canada in form of the Plaskett Research
Fellowship. Based on observations made with the NASA/ESA Hubble Space
Telescope, and obtained from the Hubble Multimission Archive, which is a
collaboration between the Space Telescope Science Institute (STScI/NASA),
the Space Telescope European Coordinating Facility (ST-ECF/ESA) and the
Canadian Astronomy Data Centre (CADC/NRC/CSA).

The authors are very thankful to the referee for the careful and thorough 
reading whose suggestions improved the paper. We would also like to thank 
Holger Baumgardt for the valuable discussions.

\bibliographystyle{mn2e}
\bibliography{references}

\begin{table*}
 \centering
 \begin{minipage}{140mm}
\caption{General properties of studied dwarf galaxies.}
\label{glist}
\begin{tabular}{p{2.3cm}p{1.3cm}p{1.3cm}cp{.7cm}p{.4cm}p{1cm}p{1cm}p{1cm}p{1.05cm}p{.7cm}}
\hline\hline
ID & R.A. & Decl. & \multicolumn{2}{c}{Morph. Type\footnote{From LEDA/NED}} & D\footnote{Distance and distance modulus from \cite{Karachentsev06,Karachentsev07}} & $m-M^b$ & E$_{B-V}$ & $M_V$ & $(V-I)_0$ & $M_{HI}$\\
& (J2000.0) & (J2000.0) & \multicolumn{2}{p{0.1cm}}{T} & Mpc & mag & mag & mag & mag & $10^7M_{\odot}$ \\
(1)        &	(2)	  &	(3)	&  (4)    &   (5) &  (6) &  (7)  & (8)  & (9)   & (10) & (11) \\
\hline
\hline
\multicolumn{2}{c}{--- Cen A/M83 complex ---}		  &		&      &       &      &    &   &       &	& \\
Cen N       & 13:48:09.2 & $-$47:33:54.0 & ?		&  dSph	& 3.77	& 27.88	& 0.079 & $-$11.15	& 1.24	& --	\\
ESO\,059-01 & 07:31:19.3 & $-$68:11:10 & 10		&  IB		& 4.57	& 28.30	& 0.147& $-$14.60	& 0.78	& 8.26	\\
ESO\,137-18 & 16:20:59.3 & $-$60:29:15 & 5.0	&  SAsc	& 6.4	& 29.03	& 0.243 & $-$17.21	& 0.79	& 34.14	\\
ESO\,215-09 & 10:57:30.2 & $-$48:10:44 & 10		&  I		& 5.25	& 28.60	& 0.221 & $-$14.08	& 0.80	& 64.21	\\
ESO\,223-09 & 15:01:08.5 & $-$48:17:33 & 9.7	&  IAB	& 6.49	& 29.06	& 0.260 & $-$16.47	& 0.88	& 63.89	\\
ESO\,269-58 & 13:10:32.9 & $-$46:59:27 & 9.4	&  I		& 3.8	& 27.90 & 0.106 & $-$15.78	& 0.98	& 2.31	\\
ESO\,269-66 & 13:13:09.2 & $-$44:53:24 & $-$5	&  dE,N	& 3.82	& 27.91 	& 0.093 & $-$13.89	& 1.00	& --	\\
ESO\,274-01 & 15:14:13.5 & $-$46:48:45 & 6.6	&  Scd	& 3.09	& 27.45	& 0.257 & $-$17.47$^a$& 1.03$^a$	& 20.18	\\
ESO\,320-14 & 11:37:53.4 & $-$39:13:14 & 10		&  I		& 6.08	& 28.92	& 0.142 & $-$13.67	& 0.80	& 2.25	\\
ESO\,381-18 & 12:44:42.7 & $-$35:58:00 & 9		&  I		& 5.32	& 28.63	& 0.063 & $-$13.39	& 0.72	& 2.71	\\
ESO\,381-20 & 12:46:00.4 & $-$33:50:17 & 9.8 	&  IBm	& 5.44	& 28.68	& 0.065 & $-$14.80	& 0.63	& 15.71	\\
ESO\,384-16 & 13:57:01.6 & $-$35:20:02 & $-$5	&  dSph/Im& 4.53	& 28.28	& 0.074 & $-$13.72	& 0.87	& --	\\
ESO\,443-09 & 12:54:53.6 & $-$28:20:27 & 10		&  Im	& 5.97	& 28.88	& 0.065 & $-$12.19	& 0.74	& 1.44	\\
ESO\,444-78 & 13:36:30.8 & $-$29:14:11 & 9.9	&  Im	& 5.25	& 28.60	& 0.053 & $-$13.48	& 0.86	& 2.06	\\
HIPASS\,J1348-37 & 13:48:33.9 & $-$37:58:03& 10 &  I		& 5.75	& 28.78 & 0.077 & $-$10.80	& 0.79	& 0.78	\\
HIPASS\,J1351-47 & 13:51:22.0 & $-$47:00:00& 10 &  I		& 5.73	& 28.79	& 0.145 & $-$11.55	& 0.68	& 2.69	\\
IC\,4247 & 13:26:44.4 & $-$30:21:45 & 1.6		&  Sab	& 4.97	& 28.48	& 0.062 & $-$14.69	& 0.66	& 3.45	\\
IKN & 10:08:05.9 & $+$68:23:57 &	$-$3			&  dSph	& 3.75	& 27.87 & 0.061 & $-$11.51\footnote{Estimated from the total $B$\,magnitudes assuming average $B-V=0.45$ and $V-I=0.7$\,mag (see Sect.\ref{SB}).}	& 0.7$^c$	& --	\\
KK\,189 & 13:12:45.0 & $-$41:49:55 & $-$5		&  dE	& 4.42	& 28.23	& 0.114 & $-$11.99	& 0.92	& --	\\
KK\,196 & 13:21:47.1 & $-$45:03:48 & 9.8		&  IBm	& 3.98	& 28.00	& 0.084 & $-$10.72	& 0.71	& --	\\
KK\,197 & 13:22:01.8 & $-$42:32:08 & 10		&  Im	& 3.87	& 27.94	& 0.154 & $-$13.04	& 1.16	& 0.17	\\
KKS\,55 & 13:22:12.4 & $-$42:43:51 & $-$3		&  dSph	& 3.94	& 27.98	& 0.146 & $-$11.17	& 1.22	& --	\\
KKS\,57 & 13:41:38.1 & $+$42:34:55 & $-$3		&  I		& 3.93	& 27.97	& 0.091 & $-$10.73	& 1.08	& --	\\
NGC\,5237 & 13:37:38.9 & $-$42:50:51 & 1.4		&  dSph/I	& 3.4	& 27.66	& 0.096 & $-$15.45	& 0.91	& 3.10	\\
\multicolumn{2}{c}{--- Sculptor group ---}	&	&	&	&	&	&	&	&	&	\\
ESO\,349-31 & 00:08:13.3 & $-$34:34:42 & 10		&  IB		& 3.21	& 27.53	& 0.012 & $-$11.87	& 0.66	& 1.34	\\
NGC\,247 & 00:47:06.1 & $-$20:39:04 & 6.7		&  SABd	& 3.65	& 27.81	& 0.018 & $-$18.76$^a$& 0.85$^a$& 37.60	\\
\multicolumn{2}{c}{--- Mafei 1\,\&\,2 ---}	&	&	&	&	&	&	&	&	&	\\
KKH\,6 & 01:34:51.6 & $+$52:05:30 & 10			&  I		& 3.73	& 27.86	& 0.351 & $-$12.66	& 0.80	& 1.34	\\
\multicolumn{2}{c}{--- IC\,342 group ---}	&	&	&	&	&	&	&	&	&	\\
KKH\,37 & 06:47:46.9 & $+$80:07:26 & 10		&  I		& 3.39	& 27.65	& 0.076 & $-$12.07	& 0.80	& 0.48	\\
UGCA\,86 & 03:59:48.3 & $+$67:08:18.6 & 9.9	&  Im	& 2.96	& 27.36	& 0.942 & $-$16.13$^d$& 0.80$^d$& 48.25	\\
UGCA\,92 & 04:32:04.9 & $+$63:36:49.0 & 10		&  Im	& 3.01	& 27.39	& 0.792 & $-$14.71	& 0.51	& 7.62	\\
\multicolumn{2}{c}{--- NGC\,2903 group ---}	&	&	&	&	&	&	&	&	&	\\
D\,564-08    & 09:19:30.0 & $+$21:36:11.7 & 10	&  I		& 8.67	& 29.69	& 0.029 & $-$12.76	& 1.00	& 1.93	\\
D\,565-06    & 09:19:29.4 & $+$21:36:12 & 10		&  I		& 9.08	& 29.79	& 0.039 & $-$12.88	& 0.95	& 0.54	\\
\multicolumn{2}{c}{--- CVn I cloud ---}	&	&	&	&	&	&	&	&	&	\\
NGC\,4068 & 12:04:02.4 & $+$52:35:19 & 9.9		&  Im	& 4.31	& 28.17	& 0.022 & $-$15.25	& 0.63	& 11.22	\\
NGC\,4163 & 12:12:08.9 & $+$36:10:10 & 9.9		&  Im	& 2.96	& 27.36	& 0.020 & $-$14.21	& 0.80	& 1.42	\\
UGC\,8215 & 13:08:03.6 & $+$46:49:41 & 9.9		&  Im	& 4.55	& 28.29	& 0.010 & $-$12.51	& 0.82	& 1.84	\\
UGC\,8638 & 13:39:19.4 & $+$24:46:33 & 9.9		&  Im	& 4.27	& 28.15	& 0.013 & $-$13.69	& 0.74	& 1.17	\\
\multicolumn{2}{c}{--- Field ---}	&	&	&	&	&	&	&	&	&	\\
D\,634-03    & 09:08:53.5 & $+$14:34:55 & 10		&  I		& 9.46	& 29.90	& 0.038  & $-$11.94	& 0.92	& 0.49	\\
DDO\,52     & 08:28:28.5 & $+$41:51:24 & 10		&  I		& 10.28	& 30.06	& 0.037 & $-$14.98	& 0.80	& 19.99	\\
ESO\,121-20 & 06:15:54.5 & $-$57:43:35 & 10		&  I		& 6.05	& 28.91	& 0.040 & $-$13.64	& 0.68	& 11.49	\\
HIPASS\,J1247-77 & 12:47:32.6 & $-$77:35:01 & 10  &  Im	& 3.16	& 27.50	& 0.748 & $-$12.86	& 0.20	& 1.05	\\
HS\,117 & 10:21:25.2 & $+$71:06:58 & 10		&  I		& 3.96	& 27.99	& 0.115 & $-$11.31	& 0.91	& --	\\
IC\,4662 & 17:47:06.3 & $-$64:38:25 & 9.7		&  IBm	& 2.44	& 26.94	& 0.070 & $-$15.58	& 0.66	& 12.58	\\
KK\,182 & 13:05:02.9 & $-$40:04:58 & 10		&  I		& 5.78	& 28.81	& 0.101 & $-$13.10	& 0.63	& 4.45	\\
KK\,230 & 14:07:10.7 & $+$35:03:37 & 10		&  I		& 1.92	& 26.42	& 0.014 & $-$9.06		& 0.74	& 0.17	\\
KK\,246 & 20:03:57.4 & $-$31:40:54 & 10		&  I		& 7.83	& 29.47	& 0.296 & $-$13.77	& 0.83	& 11.92	\\
KKH\,77 & 12:14:11.3 & $+$66:04:54 & 10		&  I		& 5.42	& 28.67	& 0.019 & $-$14.58$^a$& 0.7$^c$& 4.67	\\
NGC\,4605 & 12:39:59.4 & $+$61:36:33 & 4.9		&  SBc	& 5.47	& 28.69	& 0.014 & $-$18.41$^a$& 0.7$^c$& 25.88	\\
UGC\,7369 & 12:19:38.8 & $+$29:52:59 & 7.6		&  dE/dE,N?& 11.6	& 30.32	& 0.019 & $-$16.17	& 1.03	& --	\\
VKN & 08:40:08.9 & $+$68:26:23 & $-$3			&  dSph?	& 3.4	& --	& 0.035 & $-$10.52$^c$& 0.7$^c$	& -- \\
\hline
\hline
\end{tabular}
\end{minipage}
\end{table*}

\clearpage
\newpage
\begin{deluxetable}{p{1.8cm}p{3.9cm}p{1.9cm}p{1.8cm}p{1.6cm}p{.3cm}p{.3cm}p{.3cm}}
 \centering
\tablewidth{0pt}
\tablecaption{Globular cluster candidate properties\label{table:GCCs}}
\tablehead{
\colhead{ID} & 
\colhead{RA,DEC\,(2000)} & 
\colhead{$M_V$} & 
\colhead{$(V-I)_0$} & 
\colhead{$r_{\rm h}$} & 
\colhead{$\epsilon$} & 
\colhead{r$_{\rm proj}$} & 
\colhead{r$_{\rm proj} / {\rm r}_{\rm eff}$} \\ 
\colhead{} & 
\colhead{[hh:mm:ss],[dd:mm:ss]} & 
\colhead{[mag]} & 
\colhead{[mag]} & 
\colhead{[pc]} & 
\colhead{} & 
\colhead{[kpc]} & 
\colhead{} \\
\colhead{(1)} & 
\colhead{(2)} & 
\colhead{(3)} & 
\colhead{(4)} & 
\colhead{(5)} & 
\colhead{(6)} & 
\colhead{(7)} & 
\colhead{(8)}
}
\startdata


	&	&	&	&	&	&	&	 \\
\multicolumn{8}{c}{dIrrs}\\
	&	&	&	&	&	&	&	 \\

E115-021-01  &  02:37:54.24  $-61:18:45.39$  &$   -4.79  \pm  0.05 $& $  1.473 \pm0.052 $   &$   4.23  \pm  0.26$ &  0.18  &  2.10 &   1.52 \\
E154-023-01  &  02:56:49.17  $-54:33:16.51$  &$   -4.78  \pm  0.06 $& $  1.045 \pm0.066 $   &$   2.27  \pm  0.30$ &  0.13  &  2.01 &   0.75 \\
E154-023-02  &  02:57:01.03  $-54:35:24.44$  &$   -4.78  \pm  0.06 $& $  1.129 \pm0.062 $   &$   1.24  \pm  0.25$ &  0.20  &  3.22 &   1.09 \\
E154-023-03  &  02:57:01.06  $-54:35:11.54$  &$   -4.40  \pm  0.05 $& $  1.277 \pm0.053 $   &$   2.18  \pm  0.42$ &  0.14  &  3.07 &   1.02 \\
IC1959-01    &  03:33:20.36  $-50:23:35.17$  &$   -5.43  \pm  0.05 $& $  1.459 \pm0.051 $   &$   3.57  \pm  0.32$ &  0.07  &  3.15 &   2.61 \\
IC1959-02    &  03:33:14.24  $-50:25:40.16$  &$   -6.19  \pm  0.06 $& $  1.028 \pm0.063 $   &$   6.65  \pm  0.36$ &  0.37  &  1.51 &   1.42 \\
IC1959-03    &  03:33:15.70  $-50:23:39.77$  &$   -5.42  \pm  0.05 $& $  1.539 \pm0.051 $   &$   2.77  \pm  0.35$ &  0.42  &  2.30 &   1.83 \\
IC1959-04    &  03:33:12.43  $-50:24:52.60$  &$   -9.83  \pm  0.05 $& $  0.968 \pm0.052 $   &$   2.92  \pm  0.20$ &  0.08  &  0.03 &   0.19 \\
IC1959-05    &  03:33:11.51  $-50:24:38.34$  &$   -6.54  \pm  0.06 $& $  1.096 \pm0.062 $   &$   1.79  \pm  0.23$ &  0.17  &  0.47 &   0.22 \\
IC1959-06    &  03:33:07.70  $-50:23:17.16$  &$   -7.60  \pm  0.05 $& $  0.893 \pm0.052 $   &$   1.91  \pm  0.22$ &  0.08  &  3.08 &   2.39 \\
IC1959-07    &  03:33:15.71  $-50:25:31.32$  &$   -6.11  \pm  0.06 $& $  0.780 \pm0.074 $   &$   3.77  \pm  0.28$ &  0.10  &  1.48 &   1.41 \\
IC1959-08    &  03:33:09.50  $-50:25:45.00$  &$   -4.66  \pm  0.05 $& $  0.990 \pm0.052 $   &$   1.26  \pm  0.25$ &  0.14  &  1.77 &   1.56 \\
IC1959-09    &  03:33:03.47  $-50:25:08.15$  &$   -5.07  \pm  0.06 $& $  1.077 \pm0.070 $   &$   1.45  \pm  0.26$ &  0.42  &  2.57 &   2.07 \\
KK16-01      &  01:55:23.51  $+27:57:29.12$  &$   -5.29  \pm  0.06 $& $  1.034 \pm0.062 $   &$   2.52  \pm  0.30$ &  0.22  &  1.19 &   2.49 \\
KK17-01      &  02:00:15.85  $+28:50:42.59$  &$   -4.59  \pm  0.05 $& $  1.210 \pm0.058 $   &$   1.19  \pm  0.24$ &  0.07  &  2.11 &   8.41 \\
KK27-01      &  03:21:01.00  $-66:19:13.72$  &$   -5.24  \pm  0.06 $& $  1.570 \pm0.067 $   &$   3.21  \pm  0.19$ &  0.16  &  0.25 &   0.88 \\
KK27-02      &  03:20:57.69  $-66:19:03.74$  &$   -6.17  \pm  0.06 $& $  1.183 \pm0.071 $   &$   3.18  \pm  0.19$ &  0.05  &  0.67 &   2.61 \\
KK27-03      &  03:20:50.39  $-66:19:38.16$  &$   -4.83  \pm  0.05 $& $  1.005 \pm0.053 $   &$   1.43  \pm  0.18$ &  0.21  &  1.61 &   7.05 \\
KK65-01      &  07:42:33.50  $+16:33:40.77$  &$   -6.12  \pm  0.06 $& $  1.475 \pm0.070 $   &$   1.69  \pm  0.48$ &  0.40  &  0.92 &   0.87 \\
N1311-01     &  03:20:20.05  $-52:10:14.95$  &$   -5.03  \pm  0.05 $& $  1.089 \pm0.055 $   &$   2.22  \pm  0.29$ &  0.20  &  3.39 &   2.95 \\
N1311-02     &  03:20:17.96  $-52:11:05.37$  &$   -5.70  \pm  0.05 $& $  1.381 \pm0.051 $   &$   2.03  \pm  0.25$ &  0.24  &  2.57 &   2.24 \\
N1311-03     &  03:20:04.05  $-52:11:34.39$  &$   -6.48  \pm  0.06 $& $  1.656 \pm0.064 $   &$   4.48  \pm  0.31$ &  0.23  &  1.04 &   0.71 \\
N1311-04     &  03:20:04.00  $-52:09:21.26$  &$   -6.49  \pm  0.06 $& $  0.819 \pm0.064 $   &$   8.48  \pm  0.21$ &  0.23  &  2.97 &   2.52 \\
N1311-05     &  03:20:07.86  $-52:11:11.03$  &$   -8.60  \pm  0.05 $& $  0.900 \pm0.052 $   &$   2.09  \pm  0.20$ &  0.10  &  0.12 &   0.22 \\
N1311-06     &  03:20:06.69  $-52:10:56.18$  &$   -7.33  \pm  0.06 $& $  0.795 \pm0.067 $   &$  10.32  \pm  0.18$ &  0.07  &  0.39 &   0.39 \\
N784-01      &  02:01:18.09  $+28:48:42.92$  &$   -7.54  \pm  0.06 $& $  0.956 \pm0.062 $   &$   2.93  \pm  0.20$ &  0.15  &  2.02 &   0.98 \\
N784-02      &  02:01:12.29  $+28:50:50.86$  &$   -6.10  \pm  0.06 $& $  1.553 \pm0.064 $   &$   1.20  \pm  0.24$ &  0.29  &  1.98 &   0.75 \\
N784-03      &  02:01:16.42  $+28:49:52.81$  &$   -6.65  \pm  0.05 $& $  0.866 \pm0.056 $   &$   4.54  \pm  0.24$ &  0.12  &  0.29 &   0.25 \\
N784-04      &  02:01:16.57  $+28:51:38.43$  &$   -6.40  \pm  0.06 $& $  0.717 \pm0.076 $   &$  11.55  \pm  0.25$ &  0.09  &  2.44 &   0.88 \\
N784-07      &  02:01:18.31  $+28:49:44.61$  &$   -5.72  \pm  0.06 $& $  1.202 \pm0.074 $   &$   2.60  \pm  0.21$ &  0.10  &  0.62 &   0.37 \\
N784-08      &  02:01:21.32  $+28:49:45.61$  &$   -4.39  \pm  0.06 $& $  1.127 \pm0.078 $   &$   2.66  \pm  0.34$ &  0.37  &  1.50 &   0.67 \\
N784-09      &  02:01:18.54  $+28:52:05.14$  &$   -6.62  \pm  0.06 $& $  0.752 \pm0.070 $   &$  12.85  \pm  0.26$ &  0.08  &  3.15 &   1.18 \\
U1281-01     &  01:49:38.95  $+32:35:09.73$  &$   -7.41  \pm  0.06 $& $  1.240 \pm0.063 $   &$   6.14  \pm  0.32$ &  0.07  &  2.34 &   1.53 \\
U1281-02     &  01:49:32.54  $+32:35:25.26$  &$   -7.59  \pm  0.05 $& $  0.959 \pm0.055 $   &$   2.01  \pm  0.23$ &  0.08  &  0.38 &   0.12 \\
U3755-01     &  07:13:51.95  $+10:31:42.11$  &$   -9.20  \pm  0.06 $& $  0.773 \pm0.062 $   &$   1.62  \pm  0.25$ &  0.27  &  0.92 &   0.96 \\
U3755-02     &  07:13:50.61  $+10:30:40.01$  &$   -7.28  \pm  0.06 $& $  0.986 \pm0.062 $   &$   2.16  \pm  0.28$ &  0.09  &  1.42 &   0.85 \\
U3755-03     &  07:13:50.40  $+10:31:48.33$  &$   -6.60  \pm  0.07 $& $  0.928 \pm0.075 $   &$   4.31  \pm  0.36$ &  0.25  &  1.27 &   1.26 \\
U3755-04     &  07:13:50.12  $+10:32:15.02$  &$   -8.75  \pm  0.05 $& $  1.003 \pm0.051 $   &$   1.73  \pm  0.25$ &  0.08  &  2.21 &   1.95 \\
U3755-05     &  07:13:52.78  $+10:30:42.62$  &$   -6.66  \pm  0.05 $& $  1.038 \pm0.069 $   &$   6.81  \pm  0.46$ &  0.09  &  1.41 &   0.71 \\
U3755-06     &  07:13:52.13  $+10:31:23.98$  &$   -8.24  \pm  0.06 $& $  0.870 \pm0.063 $   &$   5.97  \pm  0.28$ &  0.05  &  0.40 &   0.48 \\
U3755-07     &  07:13:52.06  $+10:31:12.28$  &$   -7.35  \pm  0.06 $& $  0.835 \pm0.067 $   &$   7.75  \pm  0.30$ &  0.16  &  0.33 &   0.18 \\
U3755-08     &  07:13:51.43  $+10:30:57.23$  &$   -8.27  \pm  0.05 $& $  0.885 \pm0.057 $   &$   6.96  \pm  0.32$ &  0.20  &  0.72 &   0.29 \\
U3755-09     &  07:13:50.24  $+10:31:10.73$  &$   -6.64  \pm  0.06 $& $  1.194 \pm0.064 $   &$   8.54  \pm  0.37$ &  0.01  &  0.72 &   0.66 \\
U3974-01     &  07:41:59.49  $+16:49:00.92$  &$   -7.21  \pm  0.06 $& $  0.840 \pm0.063 $   &$   6.60  \pm  0.35$ &  0.16  &  2.77 &   0.86 \\
U3974-02     &  07:41:58.16  $+16:48:16.12$  &$   -8.77  \pm  0.06 $& $  0.959 \pm0.062 $   &$   2.23  \pm  0.29$ &  0.11  &  1.13 &   0.44 \\
U3974-03     &  07:41:58.08  $+16:47:50.52$  &$   -7.86  \pm  0.05 $& $  0.929 \pm0.052 $   &$   3.34  \pm  0.31$ &  0.14  &  1.23 &   0.47 \\
U3974-04     &  07:41:54.87  $+16:47:54.72$  &$   -8.28  \pm  0.06 $& $  0.979 \pm0.075 $   &$   2.73  \pm  0.29$ &  0.09  &  0.91 &   0.18 \\
U3974-05     &  07:41:54.57  $+16:48:33.71$  &$   -8.72  \pm  0.06 $& $  1.321 \pm0.072 $   &$  10.43  \pm  0.27$ &  0.17  &  1.38 &   0.30 \\
U4115-01     &  07:57:03.79  $+14:22:41.01$  &$   -7.62  \pm  0.06 $& $  0.952 \pm0.062 $   &$   3.30  \pm  0.32$ &  0.08  &  1.93 &   0.91 \\
U4115-02     &  07:56:54.05  $+14:23:40.30$  &$   -5.73  \pm  0.05 $& $  1.189 \pm0.052 $   &$   4.20  \pm  0.47$ &  0.24  &  4.28 &   3.07 \\
U4115-03     &  07:56:59.20  $+14:21:53.09$  &$   -6.45  \pm  0.06 $& $  1.221 \pm0.064 $   &$   4.50  \pm  0.39$ &  0.14  &  3.68 &   2.28 \\
U4115-04     &  07:57:01.16  $+14:24:22.25$  &$   -4.95  \pm  0.07 $& $  1.251 \pm0.089 $   &$   3.26  \pm  0.54$ &  0.36  &  2.22 &   1.76 \\
U4115-05     &  07:57:03.97  $+14:22:26.30$  &$   -4.86  \pm  0.06 $& $  1.342 \pm0.069 $   &$   0.52  \pm  0.41$ &  0.77  &  2.45 &   1.25 \\
U685-01      &  01:07:26.18  $+16:40:56.84$  &$   -7.02  \pm  0.06 $& $  0.929 \pm0.062 $   &$   3.17  \pm  0.16$ &  0.00  &  1.28 &   1.88 \\
U685-02      &  01:07:26.77  $+16:40:30.71$  &$   -4.67  \pm  0.06 $& $  1.730 \pm0.062 $   &$   1.06  \pm  0.21$ &  0.29  &  1.66 &   2.39 \\
U685-03      &  01:07:25.68  $+16:40:44.19$  &$   -7.95  \pm  0.06 $& $  0.968 \pm0.068 $   &$   8.53  \pm  0.16$ &  0.07  &  1.20 &   1.75 \\
U685-04      &  01:07:23.60  $+16:41:21.88$  &$   -8.64  \pm  0.05 $& $  0.896 \pm0.052 $   &$   1.80  \pm  0.16$ &  0.12  &  0.56 &   0.88 \\
U685-05      &  01:07:24.64  $+16:40:37.05$  &$   -7.84  \pm  0.05 $& $  0.954 \pm0.053 $   &$   4.44  \pm  0.18$ &  0.07  &  1.00 &   1.40 \\
U685-06      &  01:07:22.24  $+16:41:15.14$  &$   -8.36  \pm  0.05 $& $  0.893 \pm0.051 $   &$   1.66  \pm  0.19$ &  0.10  &  0.22 &   0.36 \\
U8760-01     &  13:50:50.73  $+38:01:48.27$  &$   -4.80  \pm  0.06 $& $  1.275 \pm0.068 $   &$   5.29  \pm  0.22$ &  0.26  &  0.60 &   0.89 \\
D565-06-01   &  09:19:29.66  $+21:36:00.15$  &$   -6.10  \pm  0.07 $& $  1.583 \pm0.094 $   &$   1.19  \pm  0.41$ &  0.07  &  0.53 &   0.72 \\
D634-03-01   &  09:08:53.72  $+14:34:55.87$  &$   -7.08  \pm  0.07 $& $  1.039 \pm0.094 $   &$   5.93  \pm  0.46$ &  0.13  &  0.30 &   0.08 \\
DDO52-01     &  08:28:27.09  $+41:51:21.71$  &$   -6.62  \pm  0.07 $& $  1.004 \pm0.099 $   &$   3.34  \pm  0.42$ &  0.09  &  0.80 &   0.43 \\
DDO52-02     &  08:28:32.70  $+41:52:26.84$  &$   -7.05  \pm  0.08 $& $  0.958 \pm0.122 $   &$   6.78  \pm  0.47$ &  0.03  &  3.90 &   2.04 \\
E059-01-01   &  07:31:18.26  $-68:11:14.49$  &$   -9.89  \pm  0.06 $& $  0.907 \pm0.077 $   &$   2.35  \pm  0.15$ &  0.05  &  0.12 &   0.01 \\
E121-20-01   &  06:15:50.13  $-57:43:27.48$  &$   -6.01  \pm  0.07 $& $  1.335 \pm0.096 $   &$   2.70  \pm  0.26$ &  0.09  &  1.11 &   2.56 \\
E223-09-01   &  15:01:08.38  $-48:16:00.56$  &$   -8.13  \pm  0.06 $& $  0.992 \pm0.078 $   &$   4.53  \pm  0.23$ &  0.11  &  2.96 &   1.00 \\
E223-09-02   &  15:01:10.40  $-48:16:00.95$  &$   -7.87  \pm  0.07 $& $  0.982 \pm0.090 $   &$   3.37  \pm  0.23$ &  0.02  &  3.00 &   1.01 \\
E223-09-03   &  15:01:02.81  $-48:17:40.69$  &$   -8.00  \pm  0.06 $& $  1.019 \pm0.078 $   &$   4.03  \pm  0.25$ &  0.19  &  1.86 &   0.65 \\
E223-09-04   &  15:01:04.59  $-48:17:29.36$  &$   -7.76  \pm  0.07 $& $  0.853 \pm0.090 $   &$   5.77  \pm  0.28$ &  0.02  &  1.30 &   0.46 \\
E223-09-05   &  15:01:05.91  $-48:17:53.93$  &$   -9.14  \pm  0.06 $& $  1.072 \pm0.077 $   &$   3.30  \pm  0.23$ &  0.10  &  1.06 &   0.38 \\
E223-09-06   &  15:01:09.85  $-48:17:33.95$  &$   -9.72  \pm  0.06 $& $  0.921 \pm0.077 $   &$   3.51  \pm  0.25$ &  0.21  &  0.36 &   0.11 \\
E223-09-07   &  15:01:09.75  $-48:18:00.87$  &$   -6.84  \pm  0.07 $& $  0.952 \pm0.094 $   &$   2.17  \pm  0.23$ &  0.21  &  0.89 &   0.32 \\
E223-09-08   &  15:01:17.37  $-48:17:54.92$  &$   -6.70  \pm  0.07 $& $  0.915 \pm0.092 $   &$   2.30  \pm  0.28$ &  0.28  &  2.79 &   0.95 \\
E269-58-01   &  13:10:25.66  $-47:00:03.41$  &$   -6.33  \pm  0.07 $& $  0.773 \pm0.092 $   &$   5.41  \pm  0.33$ &  0.15  &  1.66 &   1.16 \\
E269-58-02   &  13:10:29.41  $-46:58:19.36$  &$   -7.97  \pm  0.06 $& $  0.820 \pm0.077 $   &$   2.57  \pm  0.14$ &  0.10  &  1.54 &   1.17 \\
E269-58-03   &  13:10:31.20  $-46:59:23.11$  &$   -6.86  \pm  0.06 $& $  1.127 \pm0.078 $   &$  14.98  \pm  0.13$ &  0.18  &  0.51 &   0.32 \\
E269-58-04   &  13:10:32.08  $-47:00:57.67$  &$   -7.43  \pm  0.06 $& $  0.923 \pm0.078 $   &$   2.98  \pm  0.17$ &  0.04  &  1.65 &   1.20 \\
E269-58-05   &  13:10:35.55  $-46:58:07.17$  &$   -7.20  \pm  0.06 $& $  0.918 \pm0.078 $   &$   4.34  \pm  0.13$ &  0.03  &  1.56 &   1.25 \\
E269-58-06   &  13:10:35.88  $-46:59:32.51$  &$   -6.62  \pm  0.06 $& $  1.104 \pm0.079 $   &$  18.56  \pm  4.04$ &  0.08  &  0.39 &   0.39 \\
E269-58-07   &  13:10:39.50  $-46:59:20.17$  &$   -7.23  \pm  0.06 $& $  0.944 \pm0.078 $   &$   2.90  \pm  0.15$ &  0.06  &  1.09 &   0.93 \\
E269-58-08   &  13:10:41.34  $-46:59:00.47$  &$   -6.69  \pm  0.06 $& $  0.941 \pm0.079 $   &$   7.79  \pm  6.04$ &  0.11  &  1.52 &   1.26 \\
E320-14-01   &  11:37:50.70  $-39:12:26.91$  &$   -6.16  \pm  0.07 $& $  1.471 \pm0.092 $   &$   3.97  \pm  0.29$ &  0.17  &  1.70 &   2.89 \\
E349-031-01  &  00:08:15.16  $-34:36:35.17$  &$   -5.15  \pm  0.07 $& $  1.293 \pm0.091 $   &$   0.69  \pm  0.12$ &  0.24  &  1.81 &   4.21 \\
E349-031-02  &  00:08:12.64  $-34:36:29.97$  &$   -4.39  \pm  0.07 $& $  1.229 \pm0.101 $   &$   6.45  \pm  0.18$ &  0.18  &  1.71 &   3.99 \\
E381-20-01   &  12:46:03.21  $-33:49:16.77$  &$   -5.39  \pm  0.07 $& $  1.344 \pm0.098 $   &$   4.13  \pm  0.18$ &  0.15  &  1.78 &   1.87 \\
E384-016-01  &  13:56:54.28  $-35:18:26.01$  &$   -5.93  \pm  0.07 $& $  1.160 \pm0.090 $   &$   2.14  \pm  0.15$ &  0.22  &  2.84 &   4.54 \\
E384-016-02  &  13:56:59.90  $-35:19:37.17$  &$   -5.29  \pm  0.07 $& $  1.271 \pm0.094 $   &$   1.41  \pm  0.17$ &  0.17  &  0.67 &   1.01 \\
KK197-01     &  13:21:59.81  $-42:32:06.51$  &$   -5.69  \pm  0.07 $& $  1.050 \pm0.094 $   &$   1.95  \pm  0.17$ &  0.01  &  0.46 &   1.68 \\
KK197-02     &  13:22:02.04  $-42:32:08.14$  &$   -9.83  \pm  0.06 $& $  0.932 \pm0.077 $   &$   2.95  \pm  0.13$ &  0.11  &  0.00 &   0.00 \\
KK197-03     &  13:22:02.53  $-42:32:13.82$  &$   -7.26  \pm  0.06 $& $  0.925 \pm0.078 $   &$   2.56  \pm  0.17$ &  0.07  &  0.15 &   0.53 \\
KK246-01     &  20:03:57.47  $-31:40:55.93$  &$   -6.72  \pm  0.07 $& $  0.864 \pm0.098 $   &$   4.46  \pm  0.38$ &  0.18  &  0.19 &   0.34 \\
KK246-02     &  20:03:57.07  $-31:40:58.93$  &$   -8.30  \pm  0.06 $& $  0.960 \pm0.078 $   &$   4.24  \pm  0.26$ &  0.06  &  0.16 &   0.42 \\
KKH77-01     &  12:14:08.52  $+66:05:41.69$  &$   -7.98  \pm  0.06 $& $  0.996 \pm0.077 $   &$   1.97  \pm  0.23$ &  0.14  &  0.00 &   0.00 \\
KKH77-02     &  12:14:22.72  $+66:05:38.58$  &$   -5.41  \pm  0.07 $& $  1.074 \pm0.100 $   &$   1.77  \pm  0.25$ &  0.11  &  2.27 &   3.01 \\
KKH77-03     &  12:14:18.83  $+66:04:27.69$  &$   -5.48  \pm  0.07 $& $  1.598 \pm0.096 $   &$   4.76  \pm  0.18$ &  0.19  &  2.55 &   3.39 \\
N4163-01     &  12:12:09.70  $+36:10:15.15$  &$   -9.38  \pm  0.06 $& $  0.915 \pm0.077 $   &$   1.45  \pm  0.10$ &  0.09  &  0.17 &   0.27 \\
N4163-02     &  12:12:08.57  $+36:10:25.95$  &$   -7.83  \pm  0.06 $& $  0.969 \pm0.077 $   &$   2.08  \pm  0.10$ &  0.03  &  0.31 &   0.53 \\
N5237-01     &  13:37:37.87  $-42:51:20.02$  &$   -8.46  \pm  0.06 $& $  0.925 \pm0.077 $   &$   0.68  \pm  0.16$ &  0.39  &  0.55 &   0.57 \\
N5237-02     &  13:37:37.95  $-42:50:23.32$  &$   -6.85  \pm  0.06 $& $  0.997 \pm0.078 $   &$  15.09  \pm  0.18$ &  0.12  &  0.46 &   0.51 \\
N5237-03     &  13:37:34.62  $-42:50:01.90$  &$   -6.42  \pm  0.06 $& $  1.131 \pm0.078 $   &$   3.38  \pm  0.11$ &  0.06  &  1.11 &   1.20 \\
U8638-01     &  13:39:26.80  $+24:45:50.46$  &$   -6.57  \pm  0.06 $& $  0.871 \pm0.078 $   &$  13.63  \pm  0.88$ &  0.00  &  2.26 &   4.71 \\
U8638-02     &  13:39:24.83  $+24:46:14.04$  &$   -7.72  \pm  0.06 $& $  0.966 \pm0.077 $   &$   3.23  \pm  0.17$ &  0.05  &  1.58 &   3.25 \\
U8638-03     &  13:39:18.17  $+24:46:18.89$  &$  -10.35  \pm  0.06 $& $  1.077 \pm0.077 $   &$   2.62  \pm  0.14$ &  0.04  &  0.42 &   1.02 \\
UA86-04      &  03:59:38.20  $+67:07:12.70$  &$   -6.95  \pm  0.07 $& $  0.911 \pm0.092 $   &$   3.19  \pm  0.20$ &  0.00  &  1.42 &   5.51 \\
UA86-05      &  03:59:51.07  $+67:06:10.21$  &$   -7.53  \pm  0.07 $& $  0.878 \pm0.089 $   &$   2.19  \pm  0.11$ &  0.06  &  2.02 &   8.12 \\
UA86-07      &  03:59:45.84  $+67:07:07.54$  &$   -7.26  \pm  0.07 $& $  0.738 \pm0.091 $   &$   1.21  \pm  0.16$ &  0.13  &  1.21 &   4.57 \\
UA86-10      &  03:59:49.83  $+67:06:49.71$  &$  -11.03  \pm  0.06 $& $  0.716 \pm0.077 $   &$   1.14  \pm  0.10$ &  0.06  &  1.45 &   5.64 \\
UA86-11      &  03:59:43.02  $+67:07:27.78$  &$   -6.80  \pm  0.07 $& $  1.523 \pm0.089 $   &$   3.43  \pm  0.14$ &  0.30  &  1.02 &   3.73 \\
UA86-17      &  03:59:48.76  $+67:08:16.72$  &$   -9.67  \pm  0.06 $& $  0.731 \pm0.077 $   &$   3.27  \pm  0.13$ &  0.08  &  0.19 &   0.30 \\
UA86-20      &  03:59:42.40  $+67:08:53.83$  &$   -7.58  \pm  0.07 $& $  0.731 \pm0.089 $   &$  10.78  \pm  0.14$ &  0.19  &  0.63 &   2.95 \\
UA86-25      &  03:59:48.88  $+67:08:30.85$  &$   -7.99  \pm  0.06 $& $  0.720 \pm0.078 $   &$   4.24  \pm  0.10$ &  0.03  &  0.01 &   0.72 \\
UA86-27      &  04:00:00.75  $+67:07:37.21$  &$   -8.29  \pm  0.07 $& $  0.703 \pm0.088 $   &$   7.02  \pm  0.18$ &  0.15  &  1.26 &   5.28 \\
UA86-28      &  03:59:49.23  $+67:08:40.76$  &$   -7.79  \pm  0.07 $& $  0.868 \pm0.088 $   &$   3.11  \pm  0.15$ &  0.02  &  0.16 &   1.35 \\
UA86-29      &  03:59:50.29  $+67:08:38.16$  &$  -11.16  \pm  0.06 $& $  1.020 \pm0.077 $   &$   4.73  \pm  0.14$ &  0.12  &  0.17 &   1.37 \\
UA86-30      &  03:59:53.87  $+67:08:30.97$  &$   -7.84  \pm  0.07 $& $  0.769 \pm0.088 $   &$   4.76  \pm  0.24$ &  0.23  &  0.43 &   2.18 \\
UA92-02      &  04:32:03.24  $+63:37:06.65$  &$   -8.04  \pm  0.06 $& $  1.038 \pm0.077 $   &$   3.13  \pm  0.17$ &  0.07  &  0.28 &   0.65 \\
UA92-03      &  04:32:01.94  $+63:36:41.92$  &$   -7.74  \pm  0.06 $& $  0.784 \pm0.078 $   &$   6.20  \pm  0.12$ &  0.15  &  0.23 &   0.66 \\

	&	&	&	&	&	&	&	 \\
\multicolumn{8}{c}{dSphs}\\
	&	&	&	&	&	&	&	 \\

IKN-01       &  10:08:07.14  $+68:23:36.65$  &$   -6.65  \pm  0.06 $& $  0.911 \pm0.079 $   &$   6.62  \pm  0.29$ &  0.13  &  0.40 &   -- \\
IKN-02       &  10:08:10.79  $+68:24:05.60$  &$   -7.15  \pm  0.06 $& $  0.994 \pm0.078 $   &$   3.55  \pm  0.13$ &  0.14  &  0.50 &   -- \\
IKN-03       &  10:08:05.26  $+68:24:33.78$  &$   -6.76  \pm  0.07 $& $  1.085 \pm0.092 $   &$  14.81  \pm  0.83$ &  0.13  &  0.66 &   -- \\
IKN-04       &  10:08:04.80  $+68:24:53.71$  &$   -7.41  \pm  0.06 $& $  0.936 \pm0.077 $   &$   1.96  \pm  0.16$ &  0.18  &  1.03 &   -- \\
IKN-05       &  10:08:05.52  $+68:24:57.99$  &$   -8.47  \pm  0.06 $& $  0.906 \pm0.077 $   &$   2.89  \pm  0.13$ &  0.12  &  1.10 &   -- \\
KKS55-01     &  13:22:12.41  $-42:45:11.76$  &$   -7.36  \pm  0.06 $& $  0.907 \pm0.078 $   &$   4.51  \pm  0.18$ &  0.11  &  1.48 &   4.75 \\

	&	&	&	&	&	&	&	 \\
\multicolumn{8}{c}{dEs}\\
	&	&	&	&	&	&	&	 \\

E269-66-01   &  13:13:10.30  $-44:53:00.96$  &$   -8.08  \pm  0.06 $& $  0.911 \pm0.078 $   &$   2.87  \pm  0.15$ &  0.10  &  0.49 &   0.81 \\
E269-66-03   &  13:13:08.84  $-44:53:22.59$  &$   -9.99  \pm  0.06 $& $  0.926 \pm0.077 $   &$   2.50  \pm  0.13$ &  0.13  &  0.00 &   0.00 \\
E269-66-04   &  13:13:03.12  $-44:53:40.16$  &$   -7.18  \pm  0.06 $& $  1.017 \pm0.078 $   &$   7.48  \pm  0.13$ &  0.10  &  1.17 &   1.92 \\
E269-66-05   &  13:13:11.79  $-44:53:09.79$  &$   -7.18  \pm  0.06 $& $  0.873 \pm0.078 $   &$   4.77  \pm  0.19$ &  0.16  &  0.63 &   1.03 \\
U7369-01     &  12:19:40.78  $+29:52:04.71$  &$   -6.90  \pm  0.07 $& $  0.899 \pm0.099 $   &$   2.77  \pm  0.47$ &  0.17  &  3.43 &   2.39 \\
U7369-02     &  12:19:41.50  $+29:52:45.68$  &$   -5.83  \pm  0.09 $& $  0.846 \pm0.156 $   &$   2.89  \pm  0.53$ &  0.07  &  2.18 &   1.53 \\
U7369-03     &  12:19:39.93  $+29:52:37.20$  &$   -7.03  \pm  0.07 $& $  1.027 \pm0.102 $   &$   2.91  \pm  0.42$ &  0.03  &  1.54 &   1.07 \\
U7369-04     &  12:19:37.86  $+29:52:06.76$  &$   -5.75  \pm  0.09 $& $  0.956 \pm0.151 $   &$   6.59  \pm  0.73$ &  0.19  &  3.03 &   2.11 \\
U7369-05     &  12:19:37.32  $+29:52:08.36$  &$   -4.95  \pm  0.10 $& $  1.284 \pm0.162 $   &$   2.87  \pm  0.64$ &  0.18  &  3.04 &   2.12 \\
U7369-06     &  12:19:39.91  $+29:52:52.40$  &$   -6.65  \pm  0.07 $& $  0.996 \pm0.096 $   &$   3.16  \pm  0.53$ &  0.08  &  0.97 &   0.68 \\
U7369-09     &  12:19:38.94  $+29:53:00.92$  &$   -6.62  \pm  0.07 $& $  1.305 \pm0.098 $   &$   3.42  \pm  0.39$ &  0.41  &  0.19 &   0.14 \\
U7369-10     &  12:19:38.70  $+29:52:59.48$  &$  -12.08  \pm  0.06 $& $  0.824 \pm0.077 $   &$   2.31  \pm  0.39$ &  0.16  &  0.00 &   0.01 \\
U7369-11     &  12:19:38.88  $+29:53:05.55$  &$   -6.11  \pm  0.08 $& $  0.901 \pm0.129 $   &$   2.23  \pm  0.56$ &  0.28  &  0.37 &   0.26 \\
U7369-12     &  12:19:39.01  $+29:53:08.45$  &$   -5.89  \pm  0.08 $& $  1.152 \pm0.128 $   &$   0.55  \pm  0.39$ &  0.29  &  0.55 &   0.39 \\
U7369-13     &  12:19:40.37  $+29:53:29.93$  &$   -6.84  \pm  0.07 $& $  0.895 \pm0.101 $   &$   1.56  \pm  0.42$ &  0.18  &  2.10 &   1.47 \\
U7369-14     &  12:19:37.83  $+29:52:57.41$  &$   -6.46  \pm  0.08 $& $  1.069 \pm0.114 $   &$   3.48  \pm  0.50$ &  0.20  &  0.65 &   0.45 \\
U7369-15     &  12:19:37.45  $+29:52:52.53$  &$   -7.43  \pm  0.07 $& $  0.929 \pm0.093 $   &$   1.13  \pm  0.47$ &  0.05  &  0.99 &   0.69 \\
U7369-16     &  12:19:38.51  $+29:53:09.36$  &$   -5.66  \pm  0.09 $& $  1.204 \pm0.146 $   &$   2.29  \pm  0.56$ &  0.16  &  0.57 &   0.40 \\
U7369-17     &  12:19:37.81  $+29:53:00.54$  &$   -8.22  \pm  0.06 $& $  0.895 \pm0.079 $   &$   1.33  \pm  0.39$ &  0.05  &  0.65 &   0.45 \\
U7369-18     &  12:19:36.97  $+29:52:58.99$  &$   -7.05  \pm  0.07 $& $  1.081 \pm0.095 $   &$   1.14  \pm  0.39$ &  0.04  &  1.27 &   0.88 \\
U7369-19     &  12:19:38.27  $+29:53:26.03$  &$   -7.68  \pm  0.07 $& $  0.917 \pm0.094 $   &$   1.86  \pm  0.42$ &  0.15  &  1.52 &   1.07 \\
U7369-20     &  12:19:39.43  $+29:53:46.89$  &$   -6.27  \pm  0.08 $& $  1.007 \pm0.121 $   &$   1.31  \pm  0.42$ &  0.17  &  2.72 &   1.90 \\
U7369-21     &  12:19:37.42  $+29:53:16.35$  &$   -7.18  \pm  0.07 $& $  0.910 \pm0.096 $   &$   1.34  \pm  0.42$ &  0.13  &  1.33 &   0.93 \\
U7369-22     &  12:19:37.68  $+29:53:27.93$  &$   -6.46  \pm  0.08 $& $  0.940 \pm0.117 $   &$   1.54  \pm  0.44$ &  0.13  &  1.77 &   1.23 \\
U7369-23     &  12:19:36.34  $+29:53:12.58$  &$   -6.68  \pm  0.07 $& $  0.848 \pm0.104 $   &$   1.35  \pm  0.42$ &  0.20  &  1.88 &   1.31 \\

	&	&	&	&	&	&	&	 \\
\multicolumn{8}{c}{Sms}\\
	&	&	&	&	&	&	&	 \\

E137-18-01   &  16:20:56.66  $-60:29:08.15$  &$   -7.79  \pm  0.06 $& $  1.030 \pm0.078 $   &$   3.42  \pm  0.21$ &  0.18  &  0.65 &   0.31 \\
E137-18-02   &  16:21:05.09  $-60:27:50.06$  &$   -8.16  \pm  0.06 $& $  0.772 \pm0.079 $   &$   3.20  \pm  0.23$ &  0.10  &  2.96 &   1.40 \\
E137-18-03   &  16:21:00.44  $-60:29:10.65$  &$   -7.15  \pm  0.07 $& $  1.143 \pm0.090 $   &$   3.11  \pm  0.21$ &  0.18  &  0.29 &   0.14 \\
E137-18-04   &  16:21:00.42  $-60:29:43.48$  &$   -7.78  \pm  0.06 $& $  0.900 \pm0.078 $   &$   3.97  \pm  0.23$ &  0.04  &  0.91 &   0.43 \\
E137-18-05   &  16:21:02.92  $-60:29:14.23$  &$   -6.99  \pm  0.07 $& $  0.835 \pm0.092 $   &$   5.59  \pm  0.49$ &  0.05  &  0.82 &   0.39 \\
E137-18-06   &  16:21:00.98  $-60:30:04.40$  &$   -6.98  \pm  0.07 $& $  0.929 \pm0.095 $   &$   8.10  \pm  0.35$ &  0.00  &  1.57 &   0.74 \\
E137-18-07   &  16:21:11.03  $-60:28:37.99$  &$   -6.90  \pm  0.07 $& $  1.339 \pm0.092 $   &$   5.77  \pm  1.30$ &  0.14  &  2.92 &   1.39 \\
E274-01-01   &  15:14:15.65  $-46:47:31.48$  &$   -8.56  \pm  0.06 $& $  1.004 \pm0.077 $   &$   0.45  \pm  0.10$ &  0.41  &  0.73 &   0.39 \\
E274-01-02   &  15:14:12.16  $-46:48:39.53$  &$   -7.78  \pm  0.06 $& $  0.864 \pm0.077 $   &$   3.97  \pm  0.17$ &  0.04  &  0.48 &   0.26 \\
E274-01-03   &  15:14:15.32  $-46:48:09.76$  &$   -7.88  \pm  0.06 $& $  1.093 \pm0.077 $   &$   2.15  \pm  0.10$ &  0.09  &  0.18 &   0.10 \\
E274-01-04   &  15:14:16.49  $-46:48:17.19$  &$   -7.16  \pm  0.06 $& $  1.055 \pm0.078 $   &$   3.37  \pm  0.10$ &  0.09  &  0.30 &   0.16 \\
E274-01-06   &  15:14:19.07  $-46:48:00.91$  &$   -7.26  \pm  0.06 $& $  1.051 \pm0.077 $   &$   4.92  \pm  0.12$ &  0.11  &  0.75 &   0.40 \\
E274-01-07   &  15:14:18.54  $-46:48:24.34$  &$   -6.86  \pm  0.06 $& $  0.945 \pm0.078 $   &$   2.80  \pm  0.10$ &  0.11  &  0.63 &   0.33 \\
N247-01      &  00:47:09.72  $-20:37:40.25$  &$   -7.42  \pm  0.06 $& $  0.859 \pm0.078 $   &$  16.02  \pm  0.28$ &  0.05  &  1.75 &   0.31 \\
N247-02      &  00:47:11.59  $-20:38:48.58$  &$   -6.59  \pm  0.06 $& $  1.138 \pm0.080 $   &$  18.79  \pm  1.01$ &  0.27  &  1.40 &   0.25 \\
N4605-01     &  12:40:11.35  $+61:34:47.80$  &$   -5.69  \pm  0.07 $& $  1.491 \pm0.090 $   &$   1.49  \pm  0.18$ &  0.24  &  3.60 &   1.32 \\
N4605-02     &  12:40:05.08  $+61:35:40.51$  &$   -5.51  \pm  0.07 $& $  0.934 \pm0.101 $   &$   7.63  \pm  0.40$ &  0.10  &  1.76 &   0.65 \\
N4605-03     &  12:40:10.75  $+61:34:57.46$  &$   -7.14  \pm  0.06 $& $  0.974 \pm0.079 $   &$   9.64  \pm  0.40$ &  0.03  &  3.33 &   1.22 \\
N4605-04     &  12:40:06.32  $+61:35:40.27$  &$   -8.20  \pm  0.06 $& $  1.000 \pm0.077 $   &$   4.12  \pm  0.20$ &  0.18  &  1.92 &   0.70 \\
N4605-05     &  12:40:07.61  $+61:35:35.99$  &$   -6.77  \pm  0.06 $& $  1.227 \pm0.079 $   &$   7.13  \pm  0.21$ &  0.14  &  2.17 &   0.80 \\
N4605-06     &  12:40:09.02  $+61:35:26.86$  &$   -7.13  \pm  0.06 $& $  1.036 \pm0.079 $   &$   2.39  \pm  0.22$ &  0.00  &  2.53 &   0.93 \\
N4605-08     &  12:40:13.58  $+61:35:12.72$  &$   -6.34  \pm  0.07 $& $  0.979 \pm0.091 $   &$  13.63  \pm  0.29$ &  0.16  &  3.43 &   1.26 \\
N4605-09     &  12:40:19.99  $+61:34:24.90$  &$   -6.01  \pm  0.07 $& $  0.949 \pm0.098 $   &$   8.55  \pm  0.18$ &  0.14  &  5.17 &   1.90 \\
N4605-10     &  12:40:12.23  $+61:35:30.83$  &$   -8.26  \pm  0.06 $& $  0.969 \pm0.077 $   &$  19.16  \pm  0.25$ &  0.04  &  2.94 &   1.08 \\
N4605-11     &  12:40:16.35  $+61:35:28.57$  &$   -7.00  \pm  0.06 $& $  0.688 \pm0.080 $   &$   2.72  \pm  0.22$ &  0.13  &  3.64 &   1.33 \\
N4605-12     &  12:40:18.42  $+61:35:37.99$  &$   -5.41  \pm  0.08 $& $  1.240 \pm0.113 $   &$   2.55  \pm  0.25$ &  0.16  &  3.89 &   1.43 \\

\enddata	
\end{deluxetable}

\newpage
\begin{figure*}\setcounter{figure}{8}
\epsfig{file=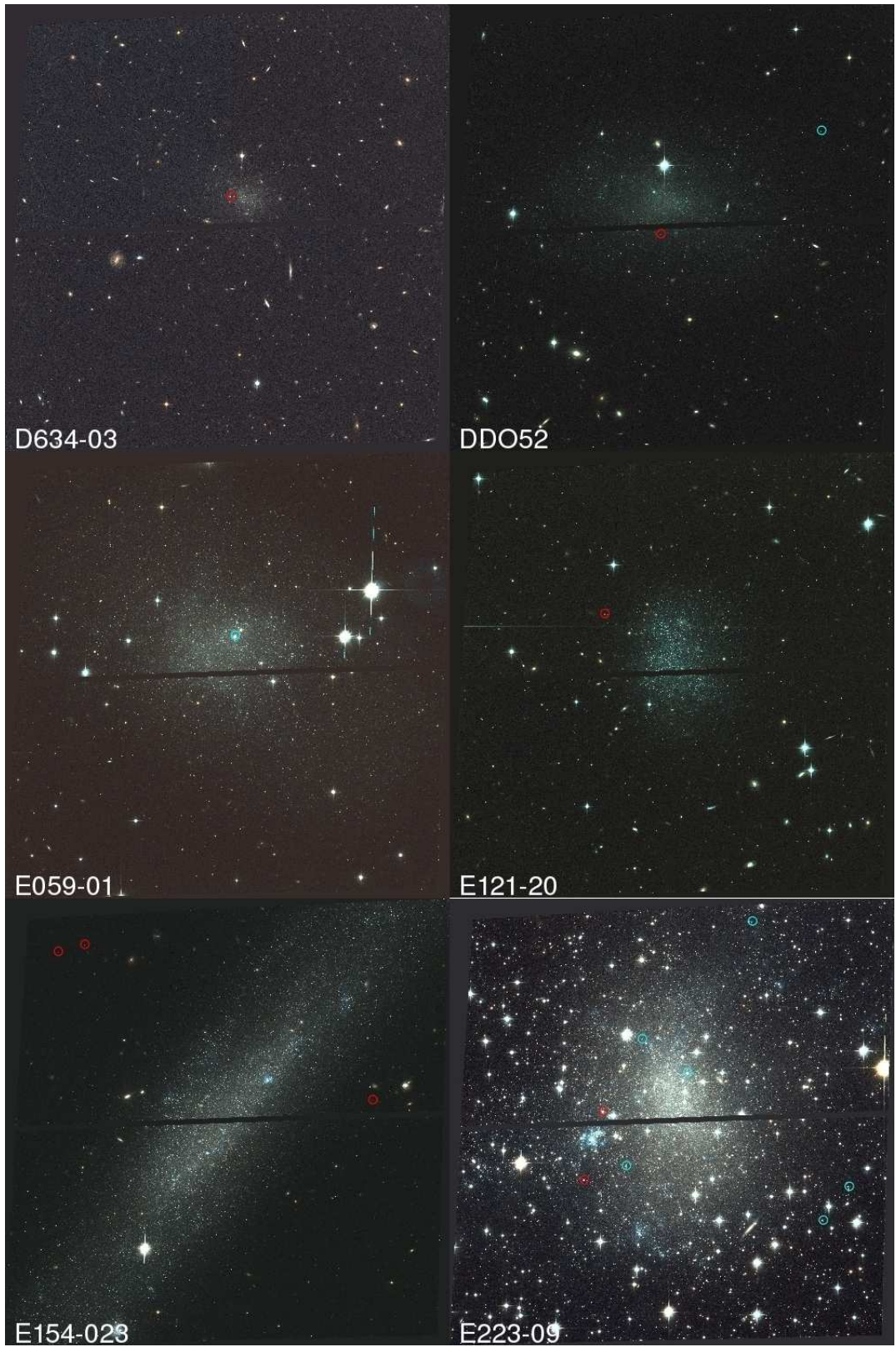, width=0.8\textwidth, bb=0 0 1000 1510}
\caption{HST/ACS color composite images of the dwarf galaxies presented 
in this study. With blue, red and magenta circles are shown the blue and 
red GC candidates and likely background contaminants, respectively. For 
the blue green and red channels we used $V, (V+I)/2$ and $I-$band HST/ACS 
images. (The full version of this figure is available upon request).
\label{color1}
}
\end{figure*}



%
%
%

\end{document}